\documentclass[a4paper,fleqn]{cas-dc}

\usepackage[numbers]{natbib}
\usepackage[figuresright]{rotating}
\DeclareUnicodeCharacter{2212}{-}
\usepackage{amsmath}
\usepackage{amssymb}
\usepackage{amsthm}
\usepackage[normalem]{ulem}
\usepackage{multirow}
\usepackage{lineno}

\newcommand{\arcsec}{\text{''}}

\def\tsc#1{\csdef{#1}{\textsc{\lowercase{#1}}\xspace}}
\tsc{WGM}
\tsc{QE}
\tsc{EP}
\tsc{PMS}
\tsc{BEC}
\tsc{DE}

\begin{document}
\let\WriteBookmarks\relax
\def\floatpagepagefraction{1}
\def\textpagefraction{.001}
\shorttitle{Radiative transfer model of IRAS4A}
\shortauthors{Bhat et~al.}

\title [mode = title]{Radiative transfer modeling of the low-mass proto-binary system, IRAS 4A1 and 4A2}                      



\author[1,2,3]{Bratati Bhat}[type=editor,
                        auid=000,bioid=1,
                        orcid=0000-0002-5224-3026]
\cormark[1]
\ead{bratati.hooghly@gmail.com}

\credit{Conceptualization of this study, Methodology, Software}


\affiliation[1]{organization={Physical Research Laboratory},
                addressline={Navrangpura},
                city={Ahmedabad},
                postcode={380009}, 
                state={Gujrat},
                country={India}}

\affiliation[2]{organization={Indian Institute of Astrophysics},
                addressline={Koramangala}, 
                city={Bangalore},
                postcode={560 034}, 
                state={Karnataka},
                country={India}}

\author[3,6]{Ankan Das}[
   ]
\author[4,5]{Prasanta Gorai}[style=chinese]
\author[1]{Dipen Sahu}[style=chinese]


\credit{Data curation, Writing - Original draft preparation}

\affiliation[3]{organization={Institute of Astronomy Space and Earth Science},
                addressline={P177, CIT Road, Scheme 7m}, 
                city={Kolkata },
                postcode={700054}, 
                state={West Bengal}, 
                country={India}}

\affiliation[4]{organization={Rosseland Centre for Solar Physics},
                addressline={University of Oslo}, 
                postcode={PO Box  1029 Blindern, 0315, Oslo}, 
                country={Norway}}

\affiliation[5]{organization={Institute of Theoretical Astrophysics},
                addressline={University of Oslo}, 
                postcode={PO Box  1029 Blindern, 0315, Oslo}, 
                country={Norway}}

\affiliation[6]{organization={Max-Planck-Institute for extraterrestrial Physics},
                addressline={P.O. Box 1312 85741}, 
                city={Garching},
                postcode={P.O. Box 1312 85741}, 
                state={Munich}, 
                country={Germany}}


\cortext[cor1]{Corresponding author}


\begin{abstract}
NGC 1333 IRAS4A is a well-studied low-mass sun-like proto-binary system. It has two components, A1 and A2, which are diverse according to their physical and chemical properties. We modeled this hot corino using the RATRAN radiative transfer code and explained different spectral signatures observed towards A1 and A2, specifically for CH$_3$OH and H$_2$CO. Our main goal is to understand the kinematical and chemical differences between A1 and A2 and to classify their dust emission and absorption properties. We considered a simple 1D spherical infalling envelope consisting of collimated outflow in the source. Recent high-resolution interferometric observations of ALMA shed new light on why the same molecular transitions towards A1 and A2 show different spectral profiles. The significant difference between spectral profiles observed towards A1 and A2 is mainly due to the dust opacity effect. Dust continuum emission toward A1 is optically thick, causing the transitions observed in absorption. Meanwhile, A2 is optically thin, leading to the observed emission profiles, and an inverse P-Cygni profile suggests the presence of an infalling envelope. Using high-resolution observations from ALMA and VLA, we expanded our model from the millimeter wavelength range to the centimeter wavelength range. This expansion demonstrates the opacity effect, which is reduced in the centimeter wavelength range, causing us to observe the lines in emission. Using our model, we reproduced the population inversion causing maser emission of methanol 44 GHz and 95 GHz transitions.

\end{abstract}




\maketitle

\section{Introduction}

Numerous nitrogen- and oxygen-bearing Interstellar complex organic molecules (iCOMs) have been detected in hot cores (dense, warm inner region of a molecular cloud surrounding a high-mass young stellar object) \citep[e.g.,][]{gora20,gora21,baek22,mond23, shim20, das24} and hot corinos (the warm, dense inner region of the envelope of the low-mass young stellar object) \citep[e.g.,][]{belb18,bell20,marb18}. iCOMs are carbon-containing molecules with at least six atoms, detected in the interstellar medium, particularly in star-forming regions. Soon after the discovery of the first hot corino, IRAS 16293-2422, by \citet{cecc04}, many such objects such as, NGC 1333 IRAS4A (hereafter IRAS4A) \citep{sant15}, NGC 1333 IRAS4B \citep{cout13}, NGC 1333 IRAS2A \citep{bott07,brin09}, \citep{tafa15, step13}, and more were detected. The IRAS 4A is located at a distance of 293pc \citep{orti18} in the NGC1333 reflection nebula region in the Perseus cloud. IRAS4A is diverse in chemical composition and a well-studied source where many complex molecules are detected to date with the single dish and interferometric telescopes \citep{mare02, bott04, bhat23}. Recently \citet{quit24} reported a total of 97 different molecular species detected from the ASAI large survey using IRAM 30 m telescope. As part of the WISH survey, the Heterodyne Instrument of the Far-infrared (HIFI) at the Herschel Space Observatory found many inverse P-Cygni lines of H$_2$O towards IRAS4A \citep{kris10,kris12,mot13}. It confirms the presence of an infalling envelope in this source; however, due to the low spatial resolution of this space-based telescope, it could not resolve the two components of IRAS4A. Previously \citet{difr01} also observed inverse P-Cygni lines of H$_2$CO and N$_2$H$^+$ towards IRAS4A using the ground-based IRAM Plateau de Bure Interferometer (PdBI) telescope with an angular resolution of ${\sim 2\arcsec}$. \citet{lope17} studied the emission of complex organic molecules from IRAS4A using Atacama Large Millimeter Array (ALMA) and pleatu de Bure Interferometer (PdBI) telescope with a higher angular resolution of $\sim 0.5\arcsec$ and resolved two cores, A1 and A2. They found striking chemical differences between the two components. Many iCOMs are observed in emission towards A2, whereas only elementary molecules are observed towards A1. The ALMA observation reveals the presence of two individual cores, A1 and A2, in IRAS4A. \citet{sahu19} observed methanol lines towards A1 and A2 in absorption and emission, respectively, at 0.84 mm with the ALMA telescope. It consists of binary components IRAS4A1 and IRAS4A2, separated by 1.8$\arcsec$ \citep{des20a} ($\sim 527$ au) and originated from the same parent cloud. The A1 and A2 show a completely different molecular spectrum. \citet{des20a} observed different lines of CH$_3$OH towards the two components of IRAS4A using VLA in $1.3$ cm regime ($18-26.5$ GHz). They observed CH$_3$OH in emission towards both A1 and A2 with the differences in intensity. The main difference between the two components is the dust opacity, where A1 is optically thick, whereas A2 is optically thin. So, a different spectrum might also arise due to the difference in the dust column density.

A bipolar red and blue shifted outflow is present in this source. Recently, \citet{taqu20} analyzed the outflow present in IRAS4A by sulfuretted species i.e OCS, CS, SO, SO$_2$. They suggested the outflow present in IRAS4A1 is younger and enriched in species initially formed in interstellar ices, whereas IRAS4A2 is comparatively older and wealthier in species formed mainly by gas phase reactions \citep{des20b}. Recently, \citet{chah24} studied the morphology and kinematics of the jets and outflows towards the IRAS 4A system using three distinct tracers: SiO, H$_2$CO, and HDCO. They found that each protostar has two outflow systems, with 4A1 showcasing three outflow cavities and 4A2 presenting four. Very recently, high resolution ($0.3\arcsec$) ALMA observations in higher frequency ($\sim$ 350 GHz) regimes towards IRAS4A reveal exciting features showing significant difference of observed spectra \citep{sahu19,suyu19} towards A1 and A2 core. Furthermore, inverse P-Cygni and emission lines of CH$_3$OH and H$_2$CO are observed towards A2, whereas the same transitions are observed in absorption towards A1. This difference is mainly due to the optically thick dust emission present towards A1.\\

An asymmetric line profile is a signature associated with infalling gas, i.e., blue-shifted emission and red-shifted absorption \citep{evan99,myer00}. Various dense gas tracers such as HCO$^{+}$, HCN, H$_2$CO, CS, N$_2$H$^+$ are used to probe the infall signatures of cores and envelops of star-forming regions by observing inverse P-Cygni line profiles where the absorption is red-shifted with respect to the source velocity. Previously, many inverse P-Cygni profiles were detected towards IRAS4A \citep{fran01,mot13} and analyzed to study the infall motion. Initially, these studies treated IRAS4A as a single object. However, recent high-resolution data have revealed that it is a protobinary system. In this paper, we analyzed the high-resolution spectra of the two distinct components, A1 and A2, separately. We utilized a radiative transfer code to model each component and to explore the physical and chemical differences between the two cores. Our goal was to determine how variations in dust opacity along these two cores affect the observed line intensities.

The paper is structured as follows. In Section \ref{sec:obs}, the details of the observational data used are given. In Section \ref{sec:model}, the details of the physical condition of the A1 and A2 cores, which we have used for the radiative transfer model, are described. In Section \ref{sec:a2}, the results from the radiative transfer model for the source IRAS4A2 are presented and discussed. The same is explained in Section \ref{sec:a1} for IRAS4A1. Finally, conclusion of this work is provided in Section \ref{sec:concl}.

\section{Observations}\label{sec:obs}
 The observations of IRAS4A were conducted using ALMA under the project code 2015.1.00147.S (PI: Yu-Nung Su). ALMA's high angular resolution enabled the observation in Band 7 (350 GHz band) using its 12 m array in C40-3 and C40-5 configurations. 
 
\begin{figure}
\centering
\includegraphics[width=0.5\textwidth]{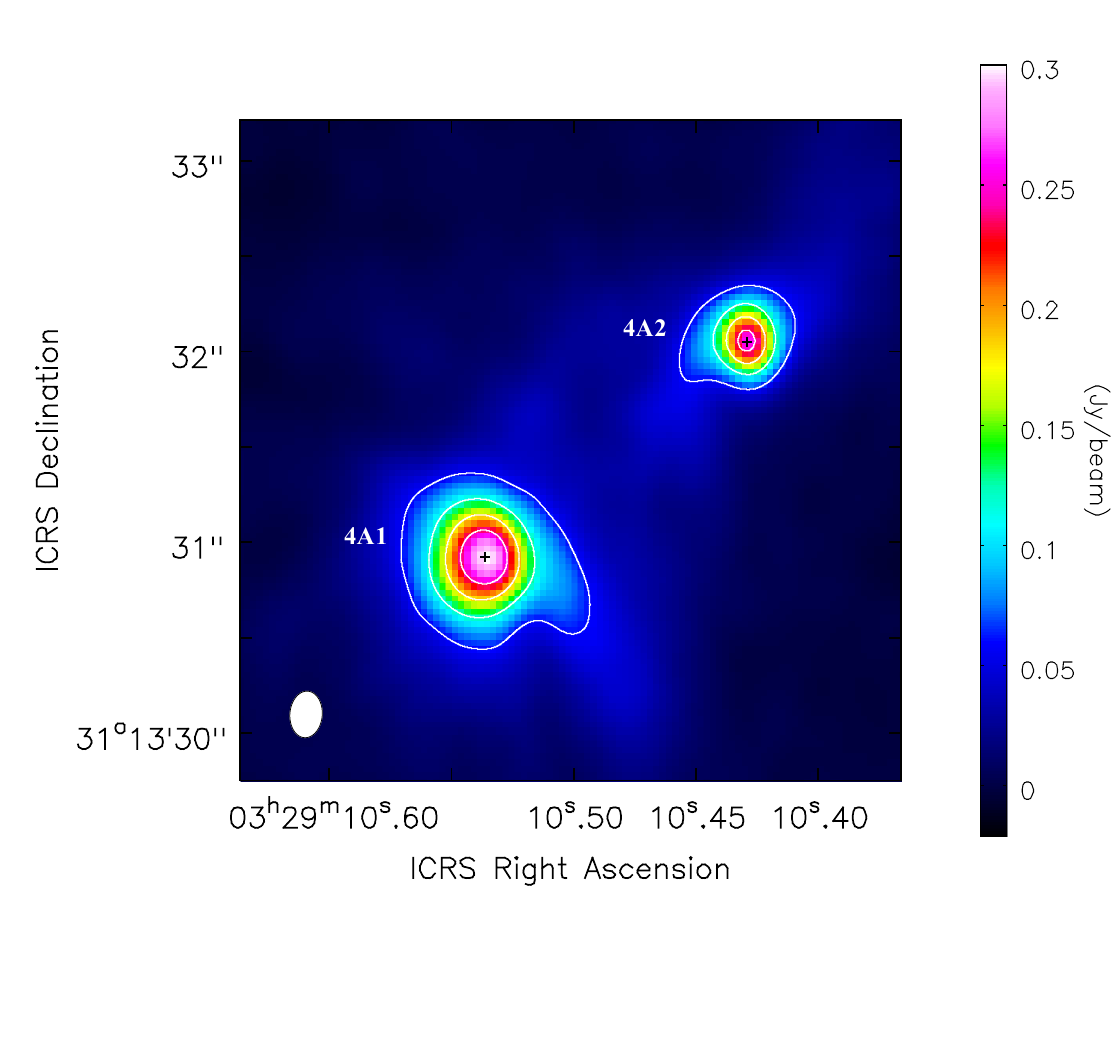}
  \caption{0.85 mm continuum image of IRAS4A system with ALMA band 7 observation at 357.45 GHz. Peak intensity of the continuum is 300.149 mJy/beam. Contour levels are at 20\%, 40\%, 60\%, and 80\% of the peak flux. The synthesized beam is shown in the lower left-hand corner of the figure. The continuum peak position for IRAS4A1 and IRAS4A2 are marked in black cross symbol. \label{fig:cont}}.
\end{figure}

  The ALMA 12m antenna primary beam's half-power width was approximately 16$\arcsec$.55 at 351.7 GHz, providing sufficient coverage of the total extent of A1 and A2. For the standard 12-m array, the maximum recoverable scale (MRS) depends on observing frequency and the array configuration. In Band 7 (350 GHz band) the maximum recoverable scale is 2$\arcsec$.444-5$\arcsec$.032 for this observation. Figure \ref{fig:cont} shows the continuum image of IRAS4A system at 0.85 mm. The resulting synthesized beam size is $0^{\arcsec}.30\times0^{\arcsec}.20$ ($\sim$ 70 au). The continuum image was generated, and the spectra were extracted using CASA (version 4.7.2). We used the calibrated data from the ALMA archive. For line identification, we used the CASSIS interactive spectrum analyzer \citep{vast15}.

We examined several emission lines of both a-CH$_3$OH and e-CH$_3$OH toward A2, which was observed in absorption toward A1. Also, we identified six transitions of H$_2$CO toward both A1 and A2, with upper state energies ranging from 50 to 240 K. \citet{sahu19} and \citet{suyu19} also reported these CH$_3$OH and H$_2$CO transitions along with some of their isotopologues using the same data. For each transition, ALMA provides a velocity resolution of 0.12 km s$^{-1}$. The observed H$_2$CO transitions cover a range of upper state energies, allowing us to trace different layers of infalling gas. \citet{suyu19} discussed the infall nature of this sources A1 and A2 on a small scale ($\sim$ 70 au) to better understand the gas kinematics. Additionally, we incorporated centimeter wavelength data for our analysis, utilizing the Very Large Array (VLA) facilities. For this, we used results obtained from \citet{des20a}.

\begin{table*}
\centering
{\scriptsize
 \caption{Molecular transitions detected towards IRAS4A system from ALMA observation \label{table:observation}.}
\begin{tabular}{|l|l|l|l|l|}
  \hline
  \hline
  Species&Frequency (GHz)&Qn NO.&E$_{up}$ (K)&log A$_{ij}$\\
  \hline
  \hline
CH$_3$OH&350.68773(e)&4(0,4) - 3(-1,3)&36.33&-4.06\\
&350.90512(a)&1(1,1) - 0(0,0)++&16.84&-3.47\\
&351.23634(e)&9(5,5) - 10(4,6)&240.51&-4.43\\
\hline
H$_2$CO&351.76864(o)&5(1,5) - 4(1,4)&62.5&-2.93\\
&362.73605(p)&5(0,5) - 4(0,4)&52.3&-2.86\\
&363.94589(p)&5(2,4) - 4(2,3)&99.5&-2.93\\
&364.10325(p)&5(4,1) - 4(4,0)&240.7&-3.30\\
&364.27514(o)&5(3,3) - 4(3,2)&158.4&-3.05\\
&364.28888(o)&5(3,2) - 4(3,1)&158.4&-3.05\\
\hline
\hline
\end{tabular}}\\
\end{table*}

\section{Models}\label{sec:model}
\subsection{Radiative transfer modeling/Physical condition}\label{sec:ratran}

 We utilized the 1D RATRAN spherically symmetric radiative transfer code \citep{hoge00} to analyze the physical properties of the IRAS4A proto-binary system. Our modeling focused on the envelope region of the IRAS4A. The primary objective of this modeling study is to comprehensively analyze the distinctions in physical, kinematical, and chemical properties between the two components of IRAS4A, A1 and A2. The study encompasses modeling various transitions of CH$_3$OH, H$_2$CO, and HCN in these two components to elucidate the observed different line shapes. \citet{mot13} and \citet{yild10} used a similar method to generate synthetic line profiles of some transitions as observed with Herschel HIFI data. However, observations at higher angular resolution using ALMA have revealed distinct differences in the spectra between A1 and A2, allowing us to retrieve variations in the physical and chemical properties between the two regions. 

\begin{figure}
\centering
\includegraphics[width=0.45\textwidth]{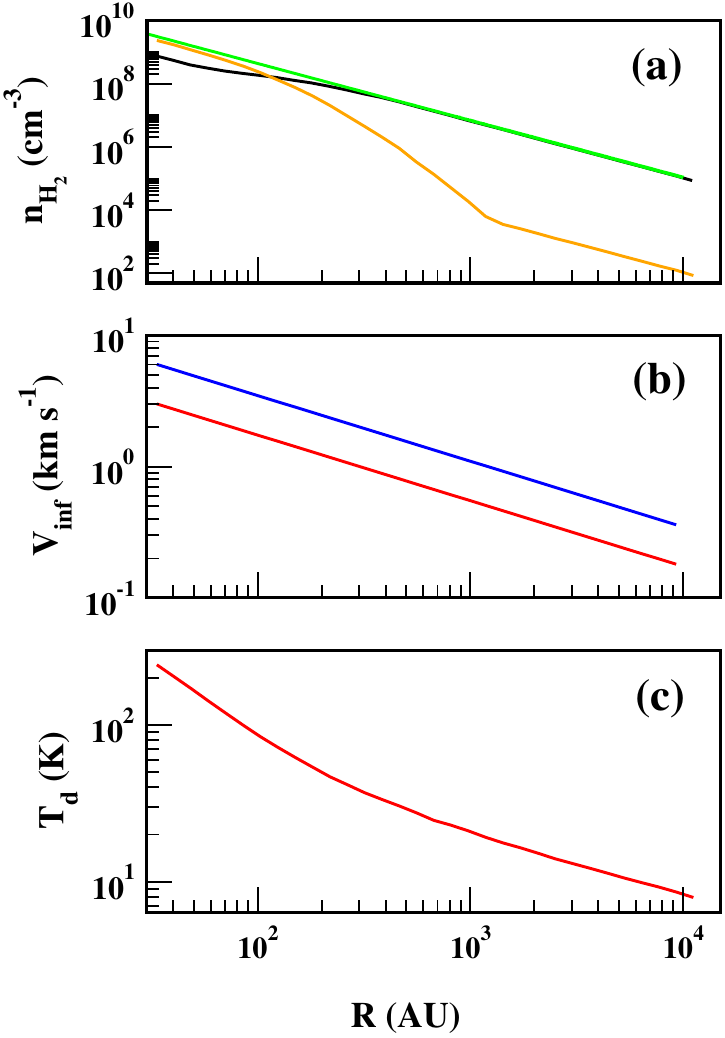}
\caption{Radial profile of the physical parameters; a. H$_2$ density (black line is for para H$_2$ and orange line is for ortho H$_2$, green line is for H$_2$ using dusty model obtained from \citet{kris12}), b. infall velocity (blue line is for V$_{1000}$=1.1 and red line is for V$_{1000}$=0.55 km s$^{-1}$.), and c. dust temperature considered here are shown.} 
  \label{fig:physical_profiles}
\end{figure}

Figure \ref{fig:physical_profiles}a illustrates the assumed spatial distribution of H$_2$ gas density in the envelope region of IRAS4A. The density profile shown in green is modeled as a power-law, with $n \propto r^{-p}$. Here, we use p=1.8, as it is determined through DUSTY model\citep[1D spherically symmetric dust radiative transfer code]{ivez97} fitting by \citet{kris12} for IRAS4A. Moreover, the inner boundary is defined at the location where the dust temperature reaches 250 K, a reasonable temperature considering the chemistry observed towards these types of sources. Material closer to the star is assumed to be located in a disk, which is not taken into account here, so the 250 K temperature is chosen, which well separates the disk region from the envelope \cite{jorg02,vank08}. The black line in Figure \ref{fig:physical_profiles}a represents para H$_2$ and the orange line ortho H$_2$. The ortho and para H$_2$ densities are calculated from the density power law assuming a thermal ortho-to-para ratio, but with a lower limit of 10$^{-3}$ \cite{paga09}. Figure \ref{fig:physical_profiles}b  displays the infall velocity structure, where the material within the envelope is assumed to be in free-fall towards the central protostar. The infall velocity (V$_{inf}$) is scaled from the velocity at 1000 au, i.e:
\begin{equation}
    V_{inf}=V_{1000}\Big(\frac{r}{1000 au}\Big)^{-0.5}.
\end{equation}
Here, we consider  V$_{1000}$=0.55 km s$^{-1}$ (red line). Additionally, we also present the infall velocity profile considering V$_{1000}$=1.1 km s$^{-1}$ (see blue line in Fig. \ref{fig:physical_profiles}b), which was used by \citet{mot13} to model the single-dish observations. 
We note that, considering a lower velocity, we find a better match between observed and modeled spectra. This discrepancy could be attributed to the low-resolution single dish data employed in \citet{mot13}, which might not have been capable of detecting the actual velocity structure in the inner region of IRAS4A where A1 and A2 are distinctly identified.
The assumed spatial temperature distribution across the envelope region is depicted in Fig. \ref{fig:physical_profiles}c. The density and temperature structure have been adopted from \citet{kris12} and \citet{mot13}. The gas and dust temperature are considered to be the same.

\begin{table}
\centering
{\scriptsize
 \caption{Key parameters used for our 1D-RATRAN modeling. \label{table:RATRAN-input-parameters}}
\begin{tabular}{ll}
  \hline
  \hline
  Input parameters & Used \\
  \hline
  \hline
 Inner radius of the envelope& $33.5$ au\\
Outer radius of envelope & $11200$ au\\
T$_{cmb}$ & $2.73$ K\\
Gas to dust mass ratio& $100$\\
 $\kappa$ (dust emissivity) & $\kappa=\kappa_0(\frac{\nu}{\nu_0})^\beta$ $^g$\\
Distance & $293$ pc\\
Bolometric temperature & $33$ K \citep{mot13}\\
  \hline
  \hline
 \end{tabular}}\\
  {\noindent $^g$ where $\kappa_0=1.06\times10^{3} $ cm$^2$/gm \citep{oss94},
 $\nu_0= 3.775 \times 10^{13}$ Hz,
 and $\beta=1 - 2$}
 \end{table} 

 The list of input parameters of our model is summarized in Table \ref{table:RATRAN-input-parameters}. It includes the inner and outer radius of the envelope region, gas-to-dust mass ratio, distance of the source, background temperature, and dust emissivity ($\kappa$). 
We use $\kappa=\kappa_0 (\frac {\nu}{\nu_0})^{\beta}$, where $\beta$ represents the power law index and $\kappa_0$ denotes the dust emissivity at a frequency $\nu_0$. We can reproduce the observed profiles for most of the transition features when $\beta=1-2$ is used in our model. This is consistent with the previously used $\beta$ = 1.5 by \citet{sahu19} and $\beta$ = 2.0 by \citet{des20a}. For modeling purposes, both para and ortho H$_2$ are considered as collision partners for radiative transfer calculations.

To compare the modeled spectra with the observed spectra, we used a beam size of 0.${\arcsec}$31 in our model (synthesized beam obtained from the observations of the CH$_3$OH and H$_2$CO lines in \citet{sahu19}) for generating the synthetic spectra. The observed spectra were extracted from 03h 29m 10.538s, +31$^{\circ}$ 13{$^\prime$} 30.93$^{\prime \prime}$
for A1 and 03h 29m 10.430s, +31$^{\circ}$ 13$^{\prime}$ 32.08$^{\prime \prime}$ for A2. We varied the modeled beam size to examine its effects and observed that this primarily affects the intensity of the simulated spectra, while the shape of the lines remains nearly unchanged. We have noted that our current 1D modeling is incapable to account for the spatial filtering inherent in interferometric data, which can miss emission on scales larger than the array's maximum recoverable size. While we tested different beam sizes to approximate this effect, this approach does not fully capture the impact of visibility sampling. As such, our analysis is most reliable for compact sources, and may underestimate the contribution from extended emissions.

\subsection{IRAS4A2}\label{sec:a2}

\subsubsection{CH$_3$OH}

\begin{figure} 
\centering
\begin{minipage}{0.40\textwidth} \includegraphics[width=\textwidth]{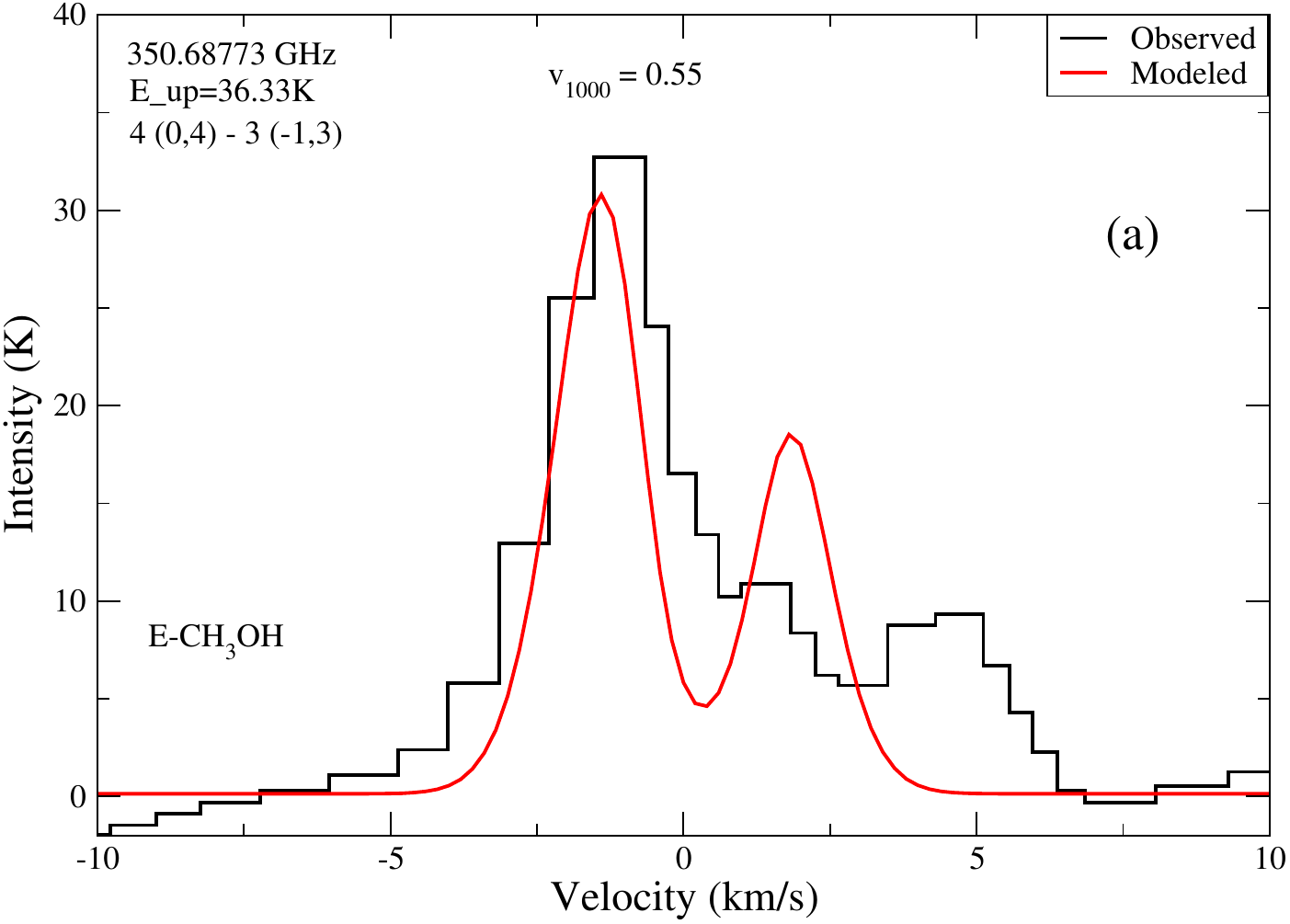} \end{minipage}
\begin{minipage}{0.40\textwidth} \includegraphics[width=\textwidth]{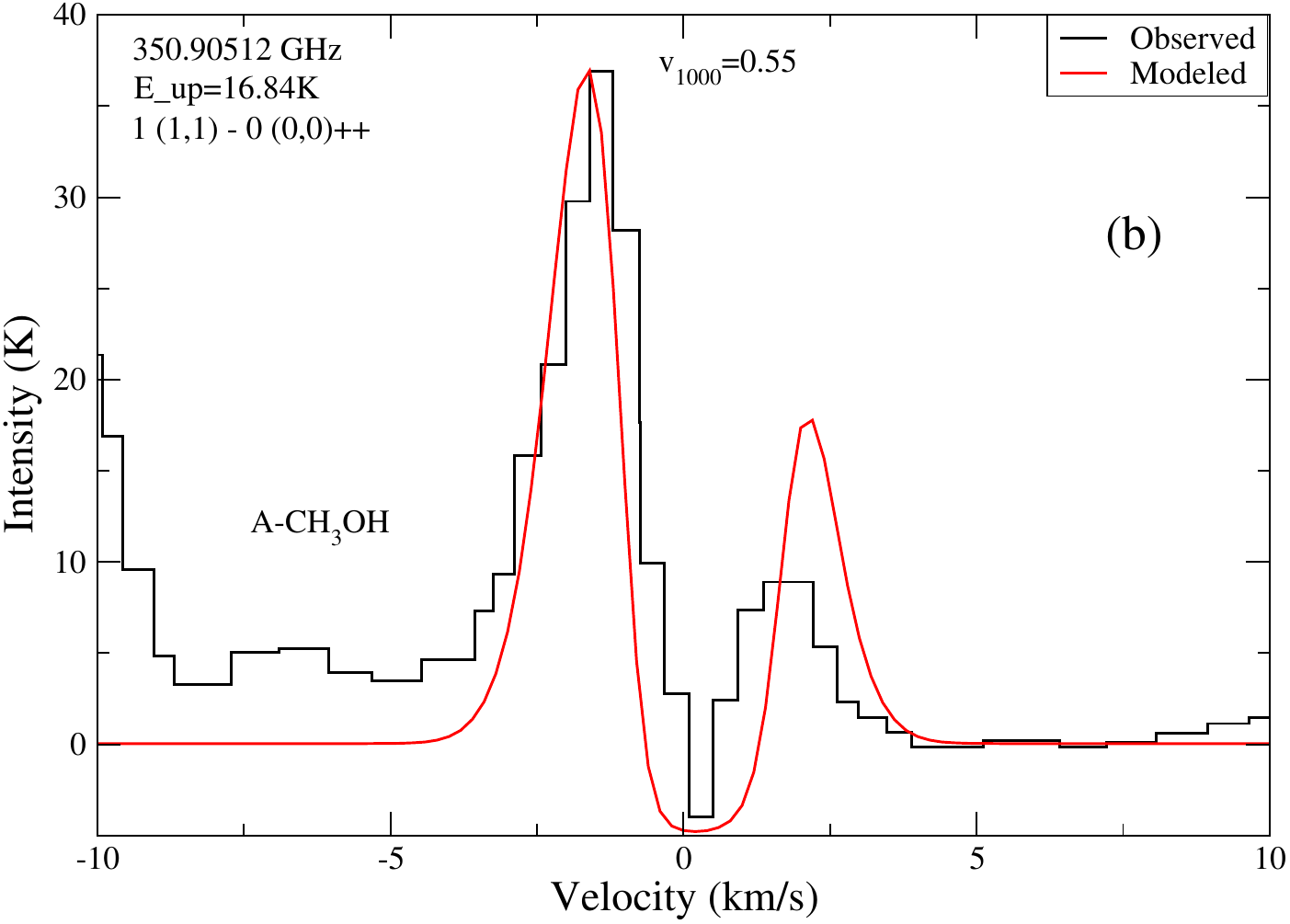} \end{minipage}
\begin{minipage}{0.40\textwidth} \includegraphics[width=\textwidth]{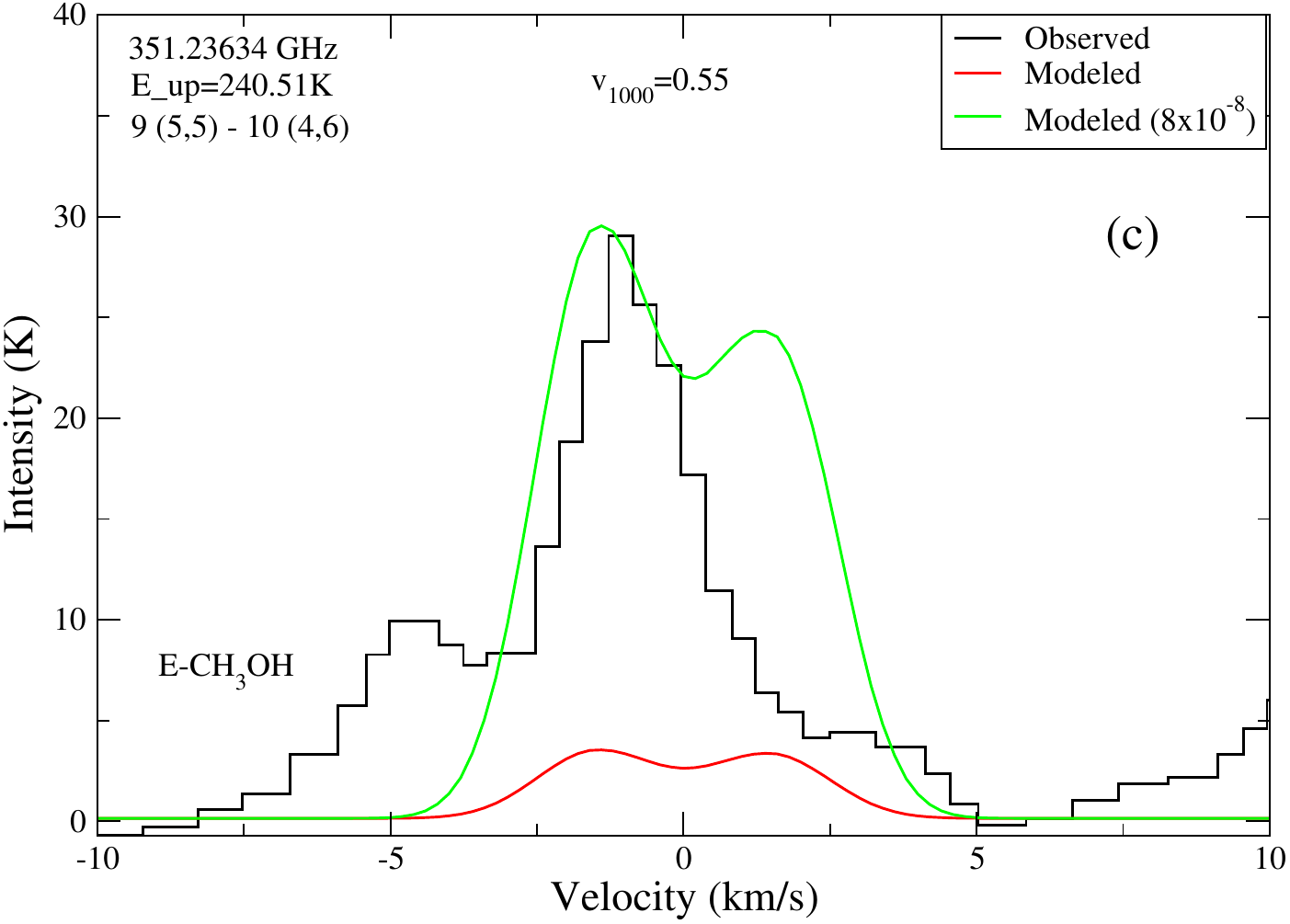} \end{minipage} 
\caption{A comparison between the observed CH$_3$OH lines (black) towards IRAS4A2 core, with modeled (red) line profiles for (a) e-CH$_3$OH 4-3 transition (b) a-CH$_3$OH 1-0 transition and (c) e-CH$_3$OH 9-10 transition using a constant abundance $8.0\times10^{-9}$ for both a-CH3OH and e-CH$_3$OH and an FWHM used is 1.5 km.s$^{-1}$ for a-CH$_3$OH and 1.67 km.s$^{-1}$ for e-CH$_3$OH. \label{fig:A2_ch3oh}}
\end{figure}

\citet{sahu19} detected CH$_3$OH along with its isotopologues: $^{13}$CH$_3$OH, CH$_2$DOH, and CH$_3$CHO in IRAS4A. With their high angular resolution of approximately 0.25$\arcsec$ ($\sim$ 70 au in linear scale), they distinguished the two components of the proto-binary system of IRAS4A.

The 350.90512 GHz transition of a-CH$_3$OH and the 350.68773 GHz and 351.23634 GHz transitions of e-CH$_3$OH are observed towards A2. An inverse P-Cygni profile is observed for the 350.90512 GHz transition ($1_{1,1} \rightarrow 0_{0,0}$) for a-CH$_3$OH, which signifies the presence of an infalling envelope in IRAS4A2. 

A radiative transfer model using RATRAN code is carried out to elucidate the observed line profiles of a-CH$_3$OH and e-CH$_3$OH. The collisional data for both a-CH$_3$OH and e-CH$_3$OH are taken from LAMDA database (\url{https://home.strw.leidenuniv.nl/~moldata/}), originally from \citet{rabl10}, with two collision partners o-H$_2$ and p-H$_2$. To explain the observed line profiles for a-CH$_3$OH and e-CH$_3$OH towards A2, we consider the dust emissivity coefficient $\kappa$ using the power law distribution mentioned in the Table \ref{table:RATRAN-input-parameters}, where, $\kappa_0$=1.06$\times10^3$ cm$^2$/gm at $\nu_0=3.775\times10^{13}$ Hz. Here, we have used the dust opacity value, $\kappa_0$, corresponding to 10$^5$ years of grain coagulation at an initial gas density of 10$^6$ cm$^{-3}$, assuming an MRN grain size distribution with thin ice mantles, as described in \cite{oss94}. $\beta=1.5$ was proposed by \citet{sahu19}. However, we find that $\beta$=1.8 better reproduces the observed spectral profile.

In Figure \ref{fig:A2_ch3oh}, we present the three observed CH$_3$OH transitions toward IRAS4 A2, along with the line profile that obtained from our radiative transfer modeling. Figure \ref{fig:A2_ch3oh}b shows the observed a-CH$_3$OH transition at a frequency of $350.9051$ GHz, while Figures \ref{fig:A2_ch3oh}a and \ref{fig:A2_ch3oh}c display the observed e-CH$_3$OH transitions at frequencies of $350.6877$ GHz and $351.2363$ GHz, respectively. Among these three CH$_3$OH lines, 350.6877 GHz transition is slightly blended with CH$_{3}$O$^{13}$CHO transitions of frequency 350.688 GHz and 350.688 GHz. The red spectra for all three transitions were generated using RATRAN modeling with a fixed abundance of $8 \times 10^{-9}$ w.r.t H$_2$ and a full width at half maximum (FWHM) of 1.5 km s$^{-1}$ for a-CH$_3$OH and 1.67 km s$^{-1}$ for e-CH$_3$OH, alongside an infall velocity $V_{1000} = 0.55$ km s$^{-1}$. These parameters were optimized based on eye estimation to achieve the best match with the observed spectra. 
 The green curve in Figure \ref{fig:A2_ch3oh}c represents RATRAN modeling with a higher abundance of $8 \times 10^{-8}$ for e-CH$_3$OH, with all other parameters held constant. The red spectra from the modeling show a reasonable match with the observed spectra (black) in Figures \ref{fig:A2_ch3oh}a and \ref{fig:A2_ch3oh}b. The transition shown in Figure \ref{fig:A2_ch3oh}a is slightly blended, leading to a minor mismatch in the peak position. In Figure \ref{fig:A2_ch3oh}c, however, the green spectrum, which has ten times higher abundance, aligns better with the observed spectra (black). The upper-state energy of the transition shown in Figure \ref{fig:A2_ch3oh}c is comparatively higher $\sim$ 240.51 K than the other two transitions ($<40$ K), indicating that it may have originated from a depper region. Despite achieving the desired intensity with the green curve in Figure \ref{fig:A2_ch3oh}c, the line profile did not align with the observation, suggesting that the density, temperature variations, and collisional rates used in the modeling do not fully reflect the conditions of the region from which it originates.

Assuming a temperature of 200 K, \citet{sahu19} estimated the average column density of CH$_3$OH to be $5.47 \times 10^{17}$ cm$^{-2}$ towards the A2. \citet{lope17} estimated the H$_2$ column density to be between $2.9 \times 10^{24}$ and $16 \times 10^{24}$ cm$^{-2}$ towards core A2 using ALMA observations at 1.2 mm. This results in a CH$_3$OH abundance ranging from $(3.42-18.8) \times 10^{-8}$ towards A2. From RATRAN modeling, we obtained a best-fit abundance of $8-80 \times 10^{-9}$ for CH$_3$OH, which is of the same order as with the values estimated from observations.

\subsubsection{\rm H$_2$CO}

\begin{figure*} 
\begin{minipage}{0.32\textwidth} \includegraphics[width=\textwidth]{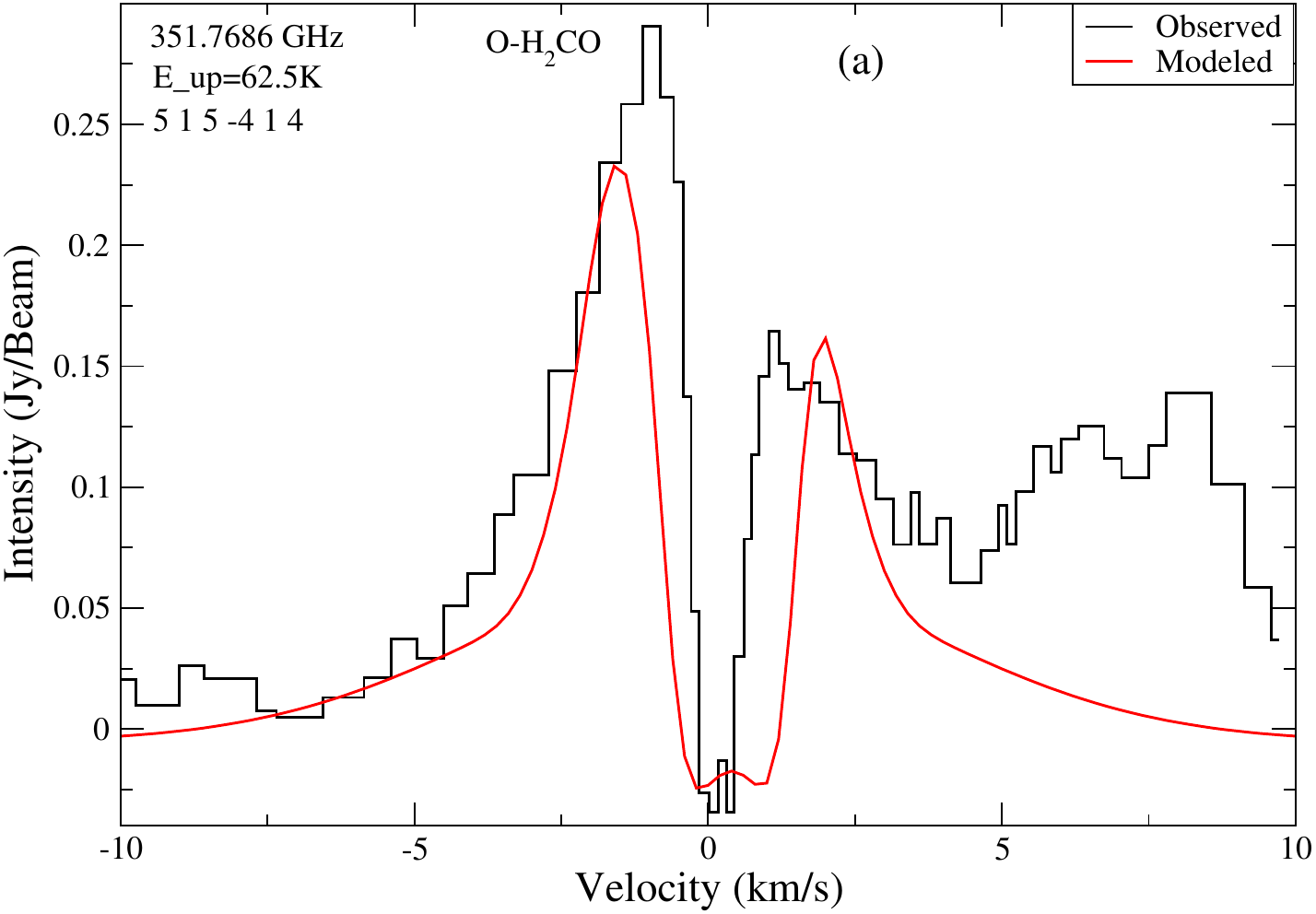} \end{minipage} 
\begin{minipage}{0.32\textwidth}
\includegraphics[width=\textwidth]{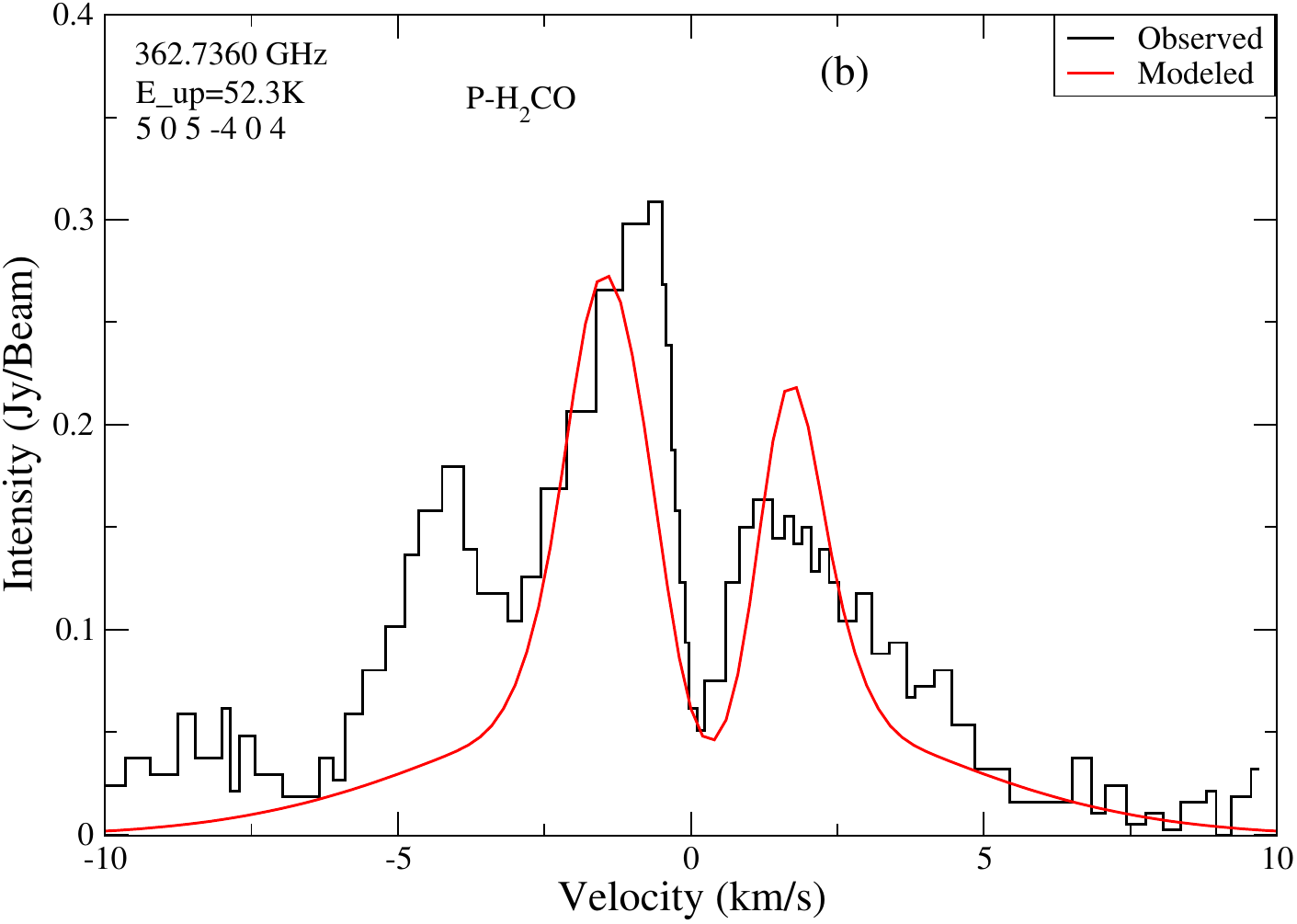} \end{minipage}
\begin{minipage}{0.32\textwidth}
\includegraphics[width=\textwidth]{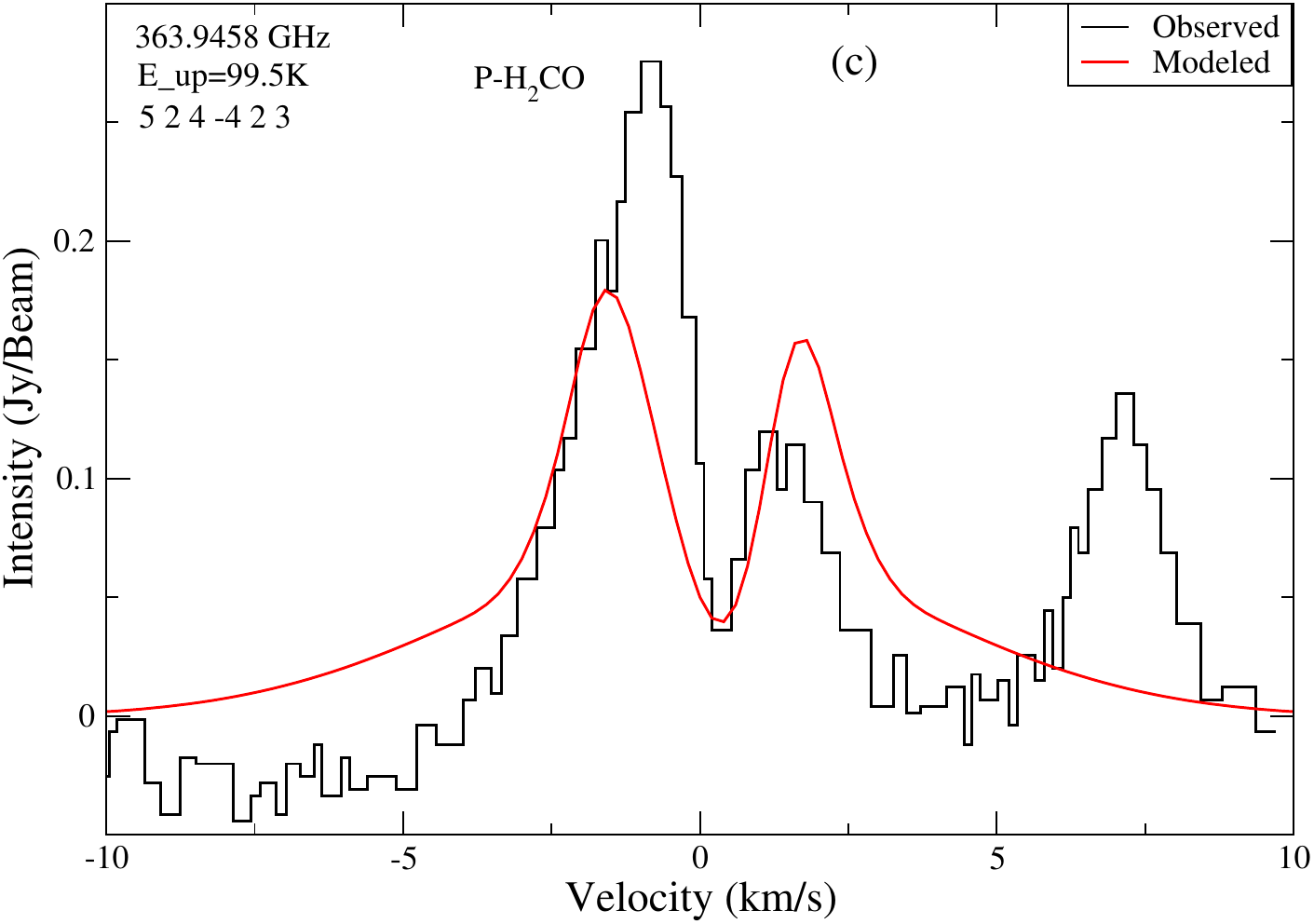} \end{minipage}
\begin{minipage}{0.32\textwidth}
\includegraphics[width=\textwidth]{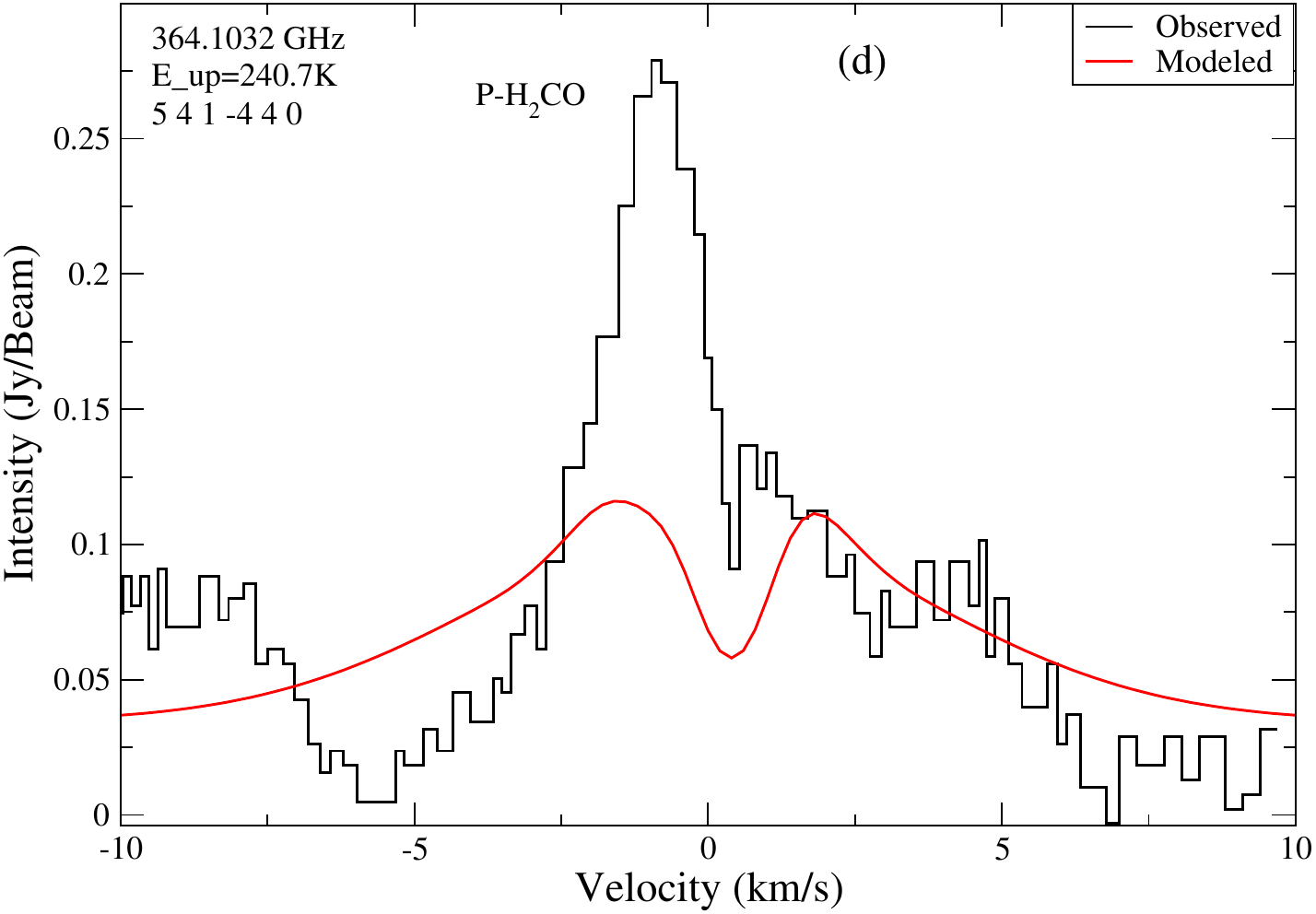} \end{minipage}
\begin{minipage}{0.32\textwidth} \includegraphics[width=\textwidth]{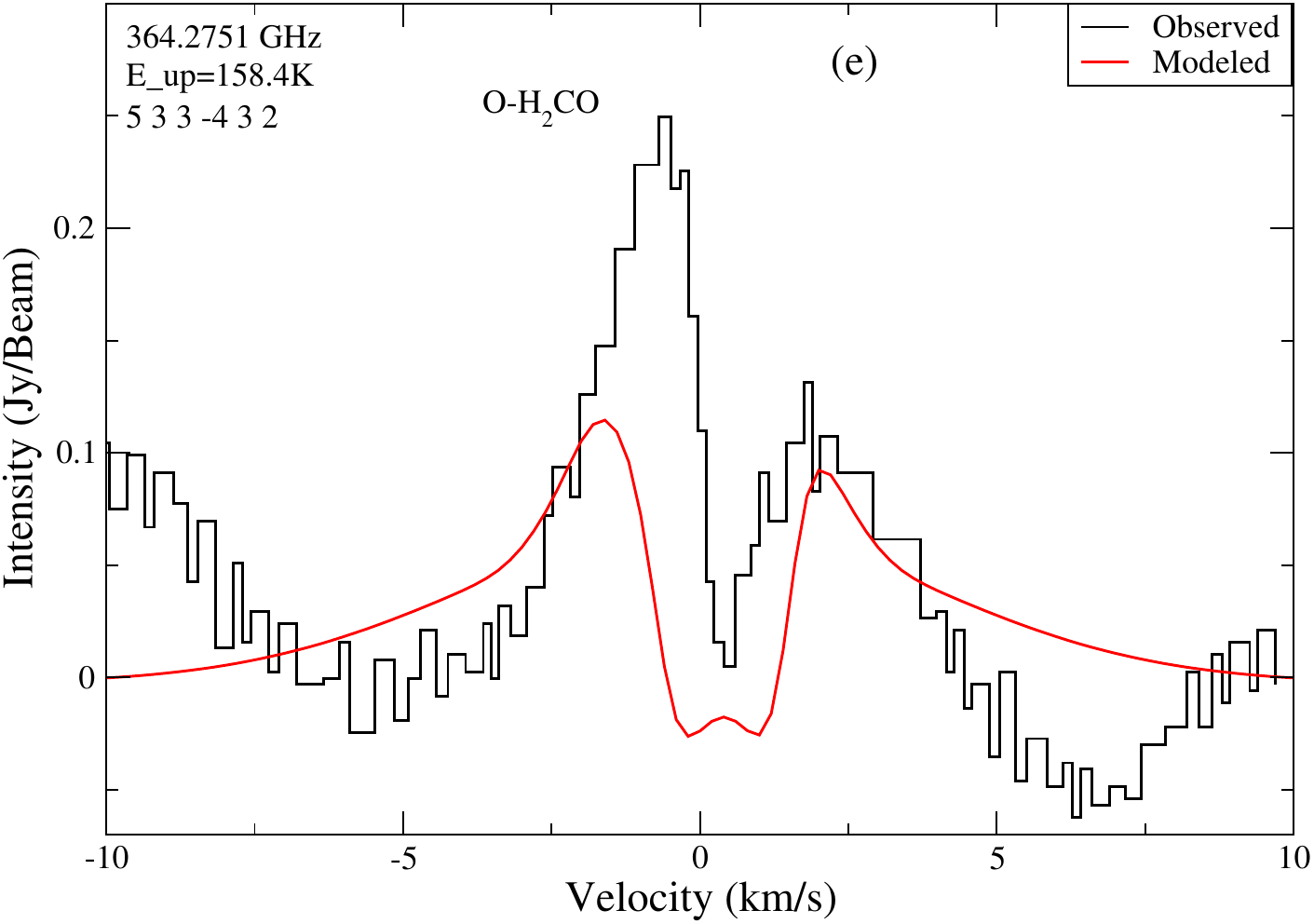} \end{minipage} 
\begin{minipage}{0.32\textwidth}
\includegraphics[width=\textwidth]{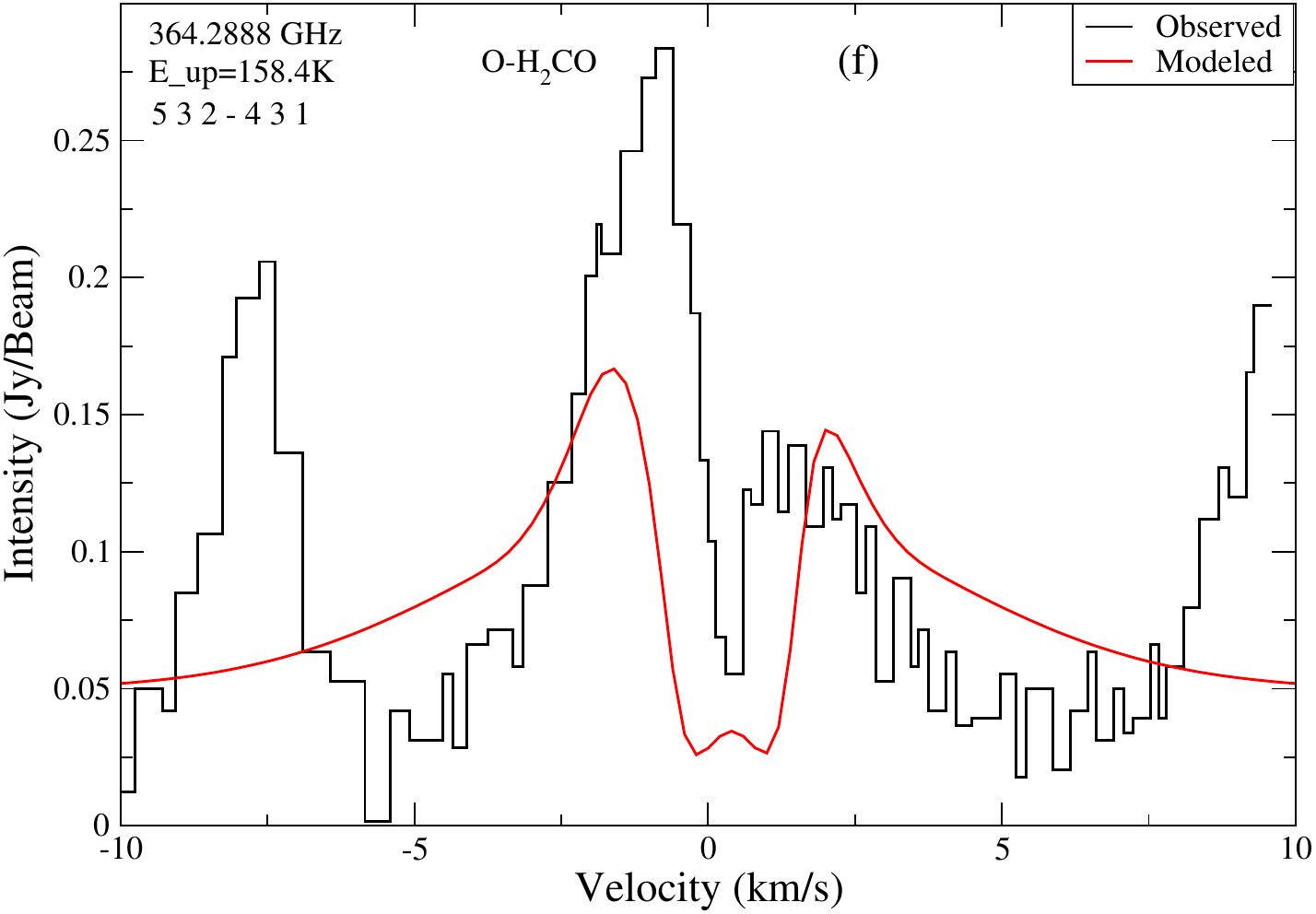} \end{minipage} 
\caption{A comparison between the observed H$_2$CO lines (black) towards IRAS4A2 core, with modeled (red) line profiles for (a) o-H$_2$CO, $5_{1,5} \rightarrow 4_{1,4}$ transition (b) p-H$_2$CO, $5_{0,5} \rightarrow 4_{0,4}$ transition (c) p-H$_2$CO, $5_{2,4} \rightarrow 4_{2,3}$ transition (d) p-H$_2$CO, $5_{4,1} \rightarrow 4_{4,0}$ transition (e) o-H$_2$CO, $5_{3,3} \rightarrow 4_{3,2}$ transition (f) o-H$_2$CO, $5_{3,2} \rightarrow 4_{3,1}$ transition using a constant abundance $1\times10^{-10}$ and a FWHM value 1.17 km s$^{-1}$. \label{fig:A2_h2co}}
\end{figure*}

 H$_2$CO is considered as a lower density gas tracer, which is useful in tracing the infalling gas material \citep{belt18}. Different transitions of H$_2$CO were observed in emission towards IRAS4A2, \citep{suyu19}. The upper state energy of the observed six transitions ranges from 52 - 241 K. The line parameters of these transitions are noted in Table \ref{table:observation}. Both o-H$_2$CO and p-H$_2$CO lines are observed towards A2. We use the radiative transfer model described in Section \ref{sec:ratran} to obtain the synthetic spectra of H$_2$CO towards IRAS4A2. The collisional data for both o-H$_2$CO and p-H$_2$CO are taken from LAMDA database (\url{https://home.strw.leidenuniv.nl/~moldata/}), originally from \citet{wies13}, with two collision partners o-H$_2$ and p-H$_2$.

  A constant abundance of $1\times10^{-10}$ for both o-H$_2$CO and p-H$_2$CO, and a FWHM of 0.83 km s$^{-1}$ (from best fit) is used in our model. A power law index $\beta$ of 1.8 is considered. \citet{suyu19} considered H$_2$CO abundance values of $2\times10^{-10}$ for r > r$_{100K}$ and $2\times10^{-8}$ for r < r$_{100K}$, where r$_{100K}$ is the distance at which the dust temperature reaches 100K toward the A2 core. The constant abundance value for H$_2$CO obtained from our modeling ($1\times10^{-10}$) is in line with the value used by \citet{suyu19} for outer envelope.

In Figure \ref{fig:A2_h2co}, all the observed lines of o-H$_2$CO and p-H$_2$CO (in black color) are plotted along with the spectra generated with our radiative transfer model (in red color) considering outflow component and foreground cloud.

A strong outflow was previously detected in the IRAS4A system. A few molecular line profiles (see Fig. \ref{fig:A2_h2co}) are significantly influenced by the outflow present in the source. \citet{suyu19} showed that in $\sim$ 10$^{\arcsec}$ angular scale, both A1 and A2 have an outflow present in the N-S direction. Also, an S-shaped morphology is observed in the outflow present in A2. \citet{des20b} reported many complex molecules present in the outflows of IRAS4A. They found a significant difference in the iCOMs present in the outflow in A1 and A2. We include an additional outflow component in the model to represent the observed H$_2$CO spectra. Our physical model initially considers no outflow; a broad Gaussian component is included during ray-tracing when the ray passes from the rear to the front half of the modeled envelope \citep{mot13,bhat22}. We observed a good match by eye estimation with the spectra when considering the outflow Gaussian component intensity of 10 K and a full width at half maximum (FWHM) of 9 km s$^{-1}$. By focusing solely on the outflow component of our radiative transfer model, we could not reproduce the absorption observed in the H$_2$CO spectra. Therefore, we added a foreground cloud in addition to the outflow that was taken into account in the line of sight toward IRAS4A2. This foreground cloud has an offset velocity of 0.4 km s$^{-1}$ relative to the source velocity obtained from RATRAN modeling. The deep absorption present in the observed spectra of H$_2$CO shown in Fig. \ref{fig:A2_h2co} can be reproduced only if the foreground cloud component is considered. This implies that H$_2$CO is absorbed by the foreground layer present in the line of sight of IRAS4A2 at an offset velocity of 0.4 km s$^{-1}$. The values of the physical input parameters used for RATRAN modeling, i.e, abundance, FWHM, are shown in Table \ref{table:RATRAN-best-fit}.



\begin{figure}  
\hskip -1.5cm
\includegraphics[width=0.5\textwidth]{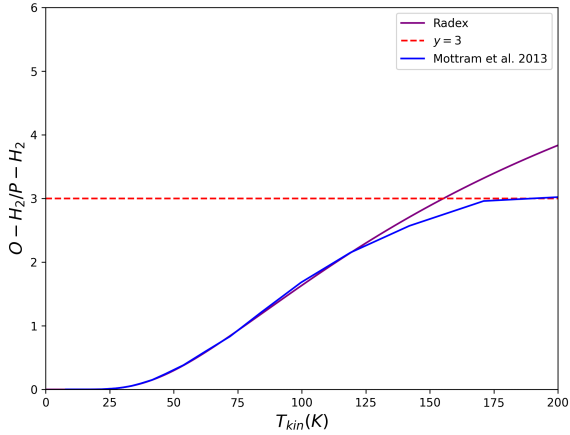}
\caption{The ratio of ortho to para H$_2$ as a function of temperature is shown. The violet line represents the ratio that is thermalized at kinetic temperature, as derived from equation \ref{eqn:opr}. The blue line indicates the ortho to para H$_2$ ratio obtained from \citet{mot13}. \label{fig:opr}}
\end{figure}

We modeled the unrelated foreground cloud separately using the Radex code \citep{vand07} to determine its density and temperature. To achieve this, we took into account both ortho and para H$_2$ as collision partners in the radiative transfer calculations. The ortho-to-para H$_2$ ratio is thermalized at the kinetic temperature according to the following relation, originally taken from \citet{flow06}: 
\begin{equation}   
\text{o-H}_2/\text{p-H}_2 \text{ or } opr = \text{minimum}(3, 9 \times e^{-170.6/T_{kin}})
\label{eqn:opr}
\end{equation}
This relation is employed in Radex \citep{van07}. We considered a total H$_2$ density of \(10^{5}\) cm\(^{-3}\), where the density of para H$_2$ is given by 
\[
\frac{n_{H_2}}{(1 + opr)} = 8.88 \times 10^{4} \text{ cm}^{-3}
\]
and the density of ortho H$_2$ is 
\[
\frac{n_{H_2}}{1 + \frac{1}{opr}} = 1.12 \times 10^{4} \text{ cm}^{-3}.
\]
A FWHM of 1.3 km s$^{-1}$ was assumed for the foreground cloud. Figure \ref{fig:opr} illustrates the variation of the ortho-para H$_2$ ratio with temperature, as derived from Radex \citep{van07} and \citet{mot13}.

We varied the foreground cloud temperature and column density of H$_2$CO between $30-60$ K and $10^{12}-10^{18}$ cm$^{-2}$, respectively.
It is noticed that the synthetic spectra for H$_2$CO best matched with the observation when cloud temperature is 40-45 K, and H$_2$CO column density of 10$^{16}$ cm$^{-2}$ is used. The column density of H$_2$ in the foreground cloud of the IRAS4A system, where H$_2$CO is absorbed, is estimated to be between $(2.9 - 16) \times 10^{24}$ cm$^{-2}$. This estimation is based on observations by \citet{quit24}, using an ALMA beam size of $0.3 \arcsec$. This corresponds to a H$_2$CO abundance of $(3.45 - 0.63) \times 10^{-9}$ in the foreground layer.

\begin{table*}
\centering
{\scriptsize
 \caption{ Abundance, linewidth, and $\beta$ obtained from our best-fitted RATRAN model are noted. \label{table:RATRAN-best-fit}.}
\begin{tabular}{|l|l|l|l|l|l|}
  \hline
  \hline
  Component&Species & \multicolumn{2}{|c|}{Abundance}& Line width &$\beta$\\
  &&This work (RATRAN)&Previous observations&(km.s$^{-1}$)&\\
  \hline
  \hline
 &a-CH$_3$OH&$8.0\times10^{-9}$&$3.41\times10^{-8}$  $^a$&1.5&1.8\\
A2 &e-CH$_3$OH&$8.0\times10^{-9}$ - $8.0\times10^{-8}$&&1.67&1.8\\
 &O-H$_2$CO&$1.0\times10^{-10}$&$2.0\times10^{-10}$  $^b$&0.83&1.8\\
 &P-H$_2$CO&$1.0\times10^{-10}$&&0.83&1.8\\
 \hline
 &a-CH$_3$OH&$9.0\times10^{-10}$&$8.76\times10^{-10}$ $^c$&1.17&1.4\\
 A1&e-CH$_3$OH&$4.0\times10^{-10}$ - $1.5\times10^{-8}$&&1.17&1.0\\
 &O-H$_2$CO&$1.1\times10^{-10}$ - $6.0\times10^{-10}$&$5.26\times10^{-11}$ $^{d}$&0.90&1.0\\
 &P-H$_2$CO&$6.0\times10^{-11}$&&0.90&1.0\\ 
  \hline
  \hline
 \end{tabular}}\\
 {\noindent $^a$\cite{sahu19} where N$_{H2}=1.6\times10^{25}$ \cite{lope17} 
 \\$^b$ \cite{suyu19}
  \\$^c$ \cite{sahu19} where N$_{H2}=1.3\times10^{26}$ \cite{sahu19} 
  \\$^d$ \cite{suyu19} where N$_{H2}=5.7\times10^{24}$ \cite{lope17}
  }
 \end{table*}

\begin{figure} 
\centering
\begin{minipage}{0.40\textwidth} 
\includegraphics[width=\textwidth]{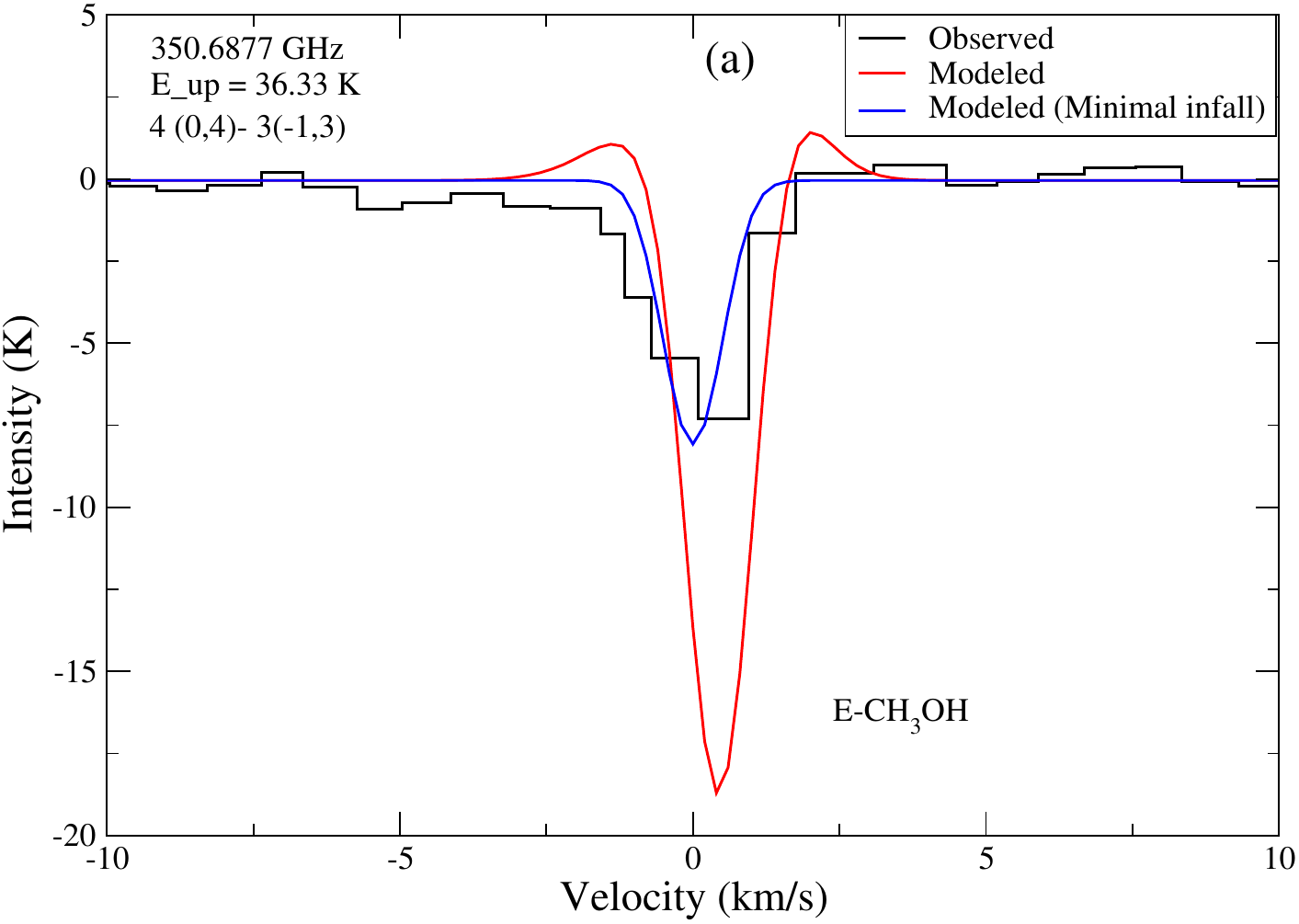} 
\end{minipage} 
\begin{minipage}{0.40\textwidth} 
\includegraphics[width=\textwidth]{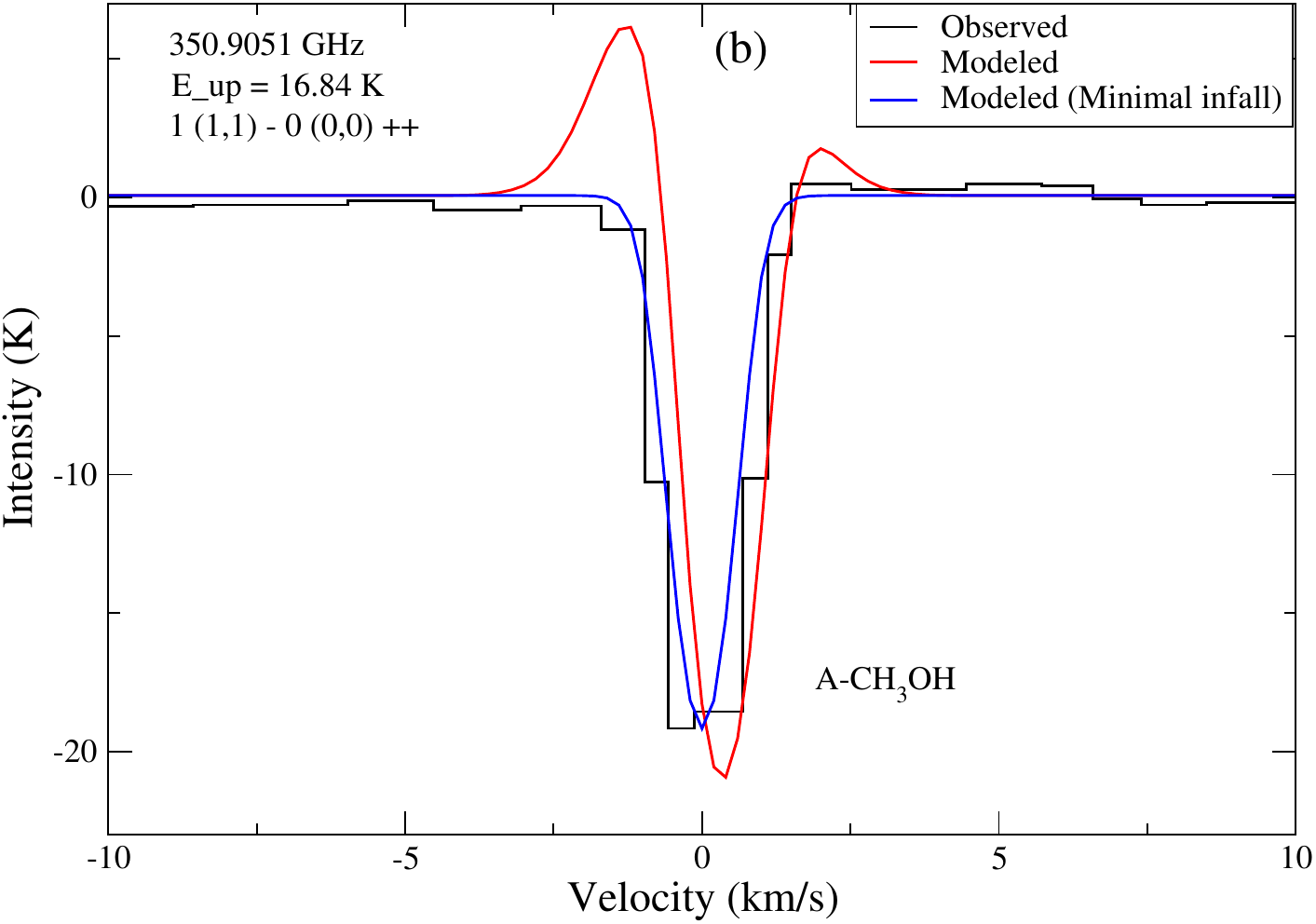} 
\end{minipage} 
\begin{minipage}{0.40\textwidth} 
\includegraphics[width=\textwidth]{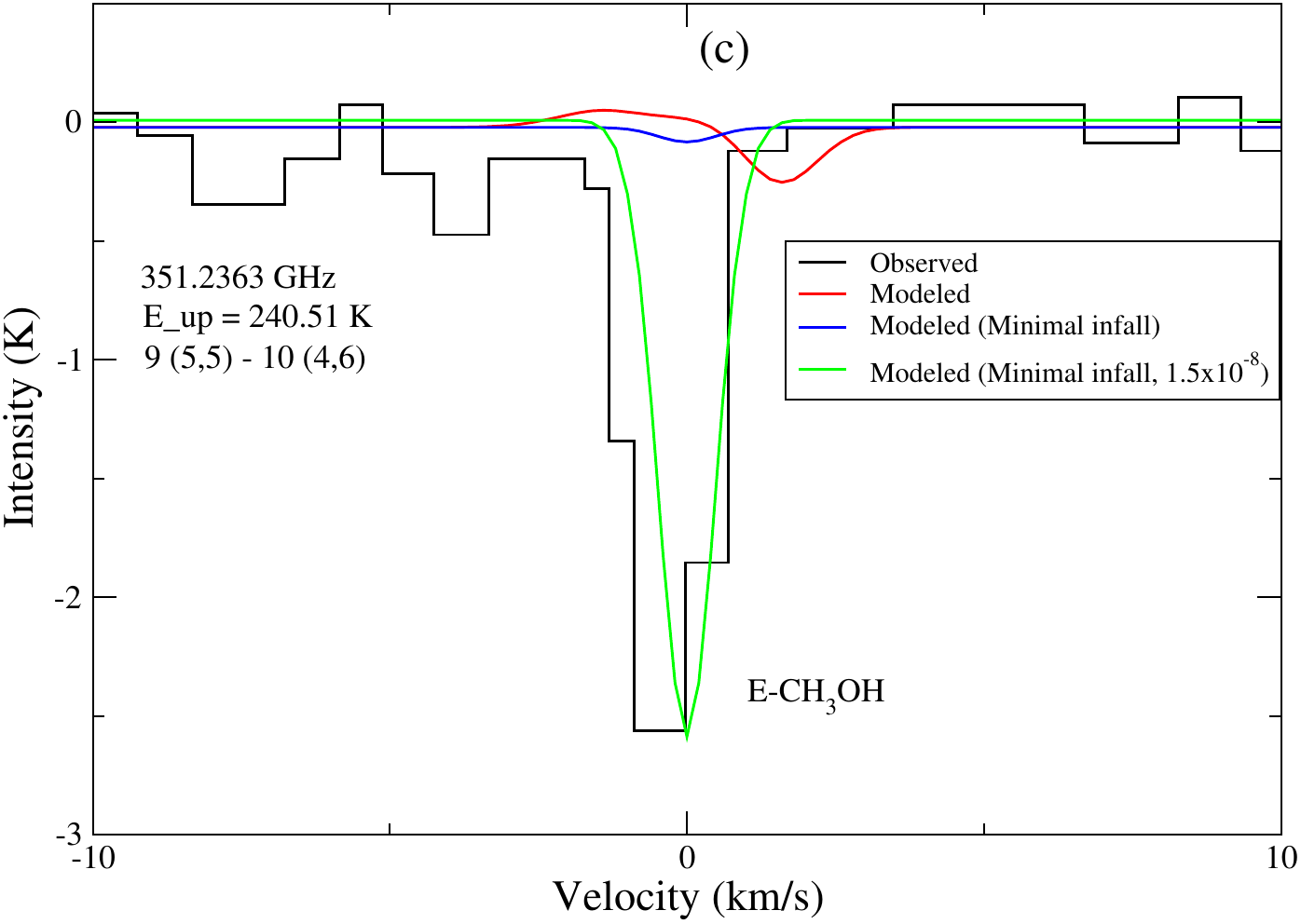} 
\end{minipage} 
\caption{A comparison between the observed CH$_3$OH lines (black) towards IRAS4A1 core, with modeled (red) line profiles for (a) e-CH$_3$OH 4-3 transition (b) a-CH$_3$OH 1-0 transition and (c) e-CH$_3$OH 9-10 transition using a constant abundance $9.0\times10^{-10}$ for a-CH$_3$OH and $4.0\times10^{-10}$ for e-CH$_3$OH. The value of FWHM used is 1.17 km.s$^{-1}$ for a-CH$_3$OH and e-CH$_3$OH. The Blue lines are modeled line profiles with the same conditions only the infall velocity considered is V$_{1000}$=0.01. \label{fig:A1_ch3oh}}
\end{figure}

\subsection{IRAS4A1}\label{sec:a1}
\subsubsection{CH$_3$OH}
The difference between the observed spectra towards A1 and A2 is investigated using radiative transfer calculations. Utilizing the model described in Section \ref{sec:ratran} of the RATRAN code, we generated synthetic spectra for the CH$_3$OH lines observed towards IRAS4A1. Notably, the primary factor responsible for the observed absorption spectra towards A1 is the dust emissivity coefficient $\kappa$. 
Thus, to model the observed CH$_3$OH lines towards A1, we considered a high dust emissivity represented by a power-law distribution, with the same $\kappa_0$ that was used for IRAS4 A2 in Section {\ref{sec:a2}} (i.e.,  $\kappa_0=1.06\times10^{3}$ cm$^{2}$/gm at $\nu_0=3.775\times10^{13}$). We used an emissivity power-law coefficient $\beta=1.4$ to best match the observations, along with a constant abundance of $9\times10^{-10}$ and a full width at half maximum (FWHM) of 1.17 km s$^{-1}$ for a-CH$_3$OH. In Figure \ref{fig:A1_ch3oh}b, the observed spectrum of a-CH$_3$OH is shown in black, while the modeled spectrum is depicted in red. The synthetic spectrum for a-CH$_3$OH (the red spectrum in panel b) is asymmetric and does not exhibit pure absorption due to the presence of an infall velocity. When we set the infall velocity to a minimal value, V$_{1000}$=0.01, with the same constant abundance of $9\times10^{-10}$, or use a nearly static envelope in the model, we obtain a pure absorption spectrum (shown as the blue spectrum in panel b) that aligns well with the observations.

For e-CH$_3$OH, we used the same constant abundance of 9$\times10^{-10}$ and an FWHM of 1.17 km s$^{-1}$ to match the observed spectra best (see Fig. \ref{fig:A1_ch3oh}a). For e-CH$_3$OH, a $\beta$ value of 1.0 was used for the best match. In Figures \ref{fig:A1_ch3oh}a and \ref{fig:A1_ch3oh}c, two observed transitions of e-CH$_3$OH are presented with black lines. Considering the previously described infall velocity structure, the synthetic spectra are shown in red. The red spectra are asymmetric and do not match well with the observed spectra, which exhibit pure absorption and symmetry characteristics. When we assume the infall velocity to be a minimal value (representing a nearly static envelope case), the modeled spectra (blue spectra in Figure \ref{fig:A1_ch3oh}a and \ref{fig:A1_ch3oh}c) align well in case of 36.33K transition (panel a) with the observed spectra (black). 

For the transition at 351.2363 GHz (panel c), the intensity of the spectra is underrepresented. Given that the upper state energy of this e-CH$_3$OH transition at 9 - 10 is relatively high (approximately 241 K) and may have originated from the inner region, a higher abundance value of about  $1.5\times10^{-8}$ is applied to achieve the desired intensity of the observed spectra (shown in green). In their analysis of the rotational diagram of the observed CH$_3$OH lines towards the region A1, \citet{sahu19} determined a CH$_3$OH column density of $1.14\times10^{17}$ cm$^{-2}$. Additionally, they reported an H$_2$ column density for the same region of $1.30\times10^{26}$ cm$^{-2}$. Using these values, the calculated abundance of CH$_3$OH is $8.8\times10^{-10}$, which is consistent with the order of magnitude ($10^{-10}$) of the abundance derived from the best-fit value obtained from the RATRAN modeling presented in Table \ref{table:RATRAN-best-fit}.

Our radiative transfer modeling suggests that the high concentration of dust in the direction of A1 is obscuring the emission lines, resulting in their appearance as absorption features. This phenomenon was previously discussed by \citet{suyu19}, who explained that the high dust opacity is responsible for the observed absorption of the lines toward A1.

\begin{figure*} 
\begin{minipage}{0.32\textwidth} \includegraphics[width=\textwidth]{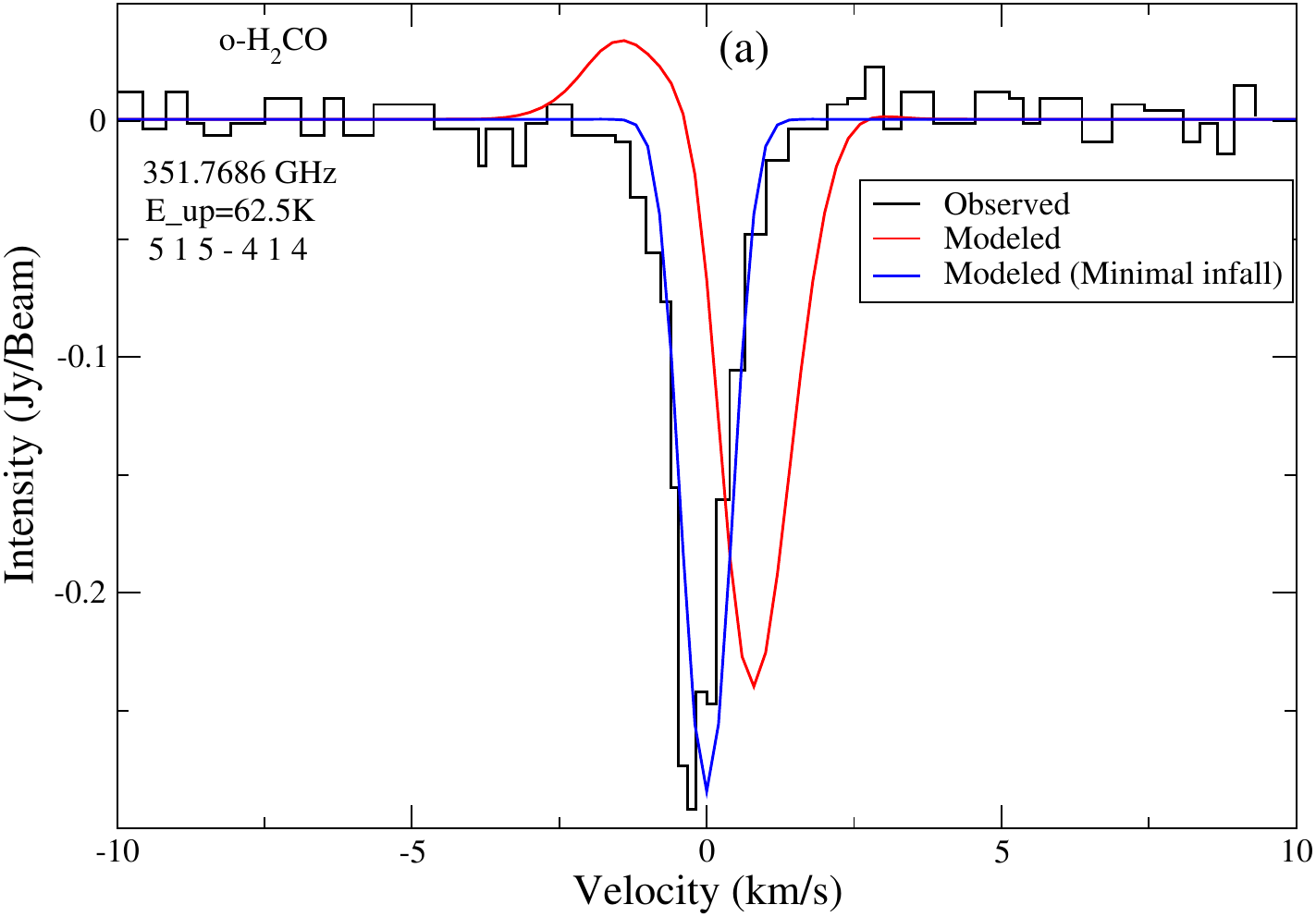} \end{minipage} 
\begin{minipage}{0.32\textwidth} \includegraphics[width=\textwidth]{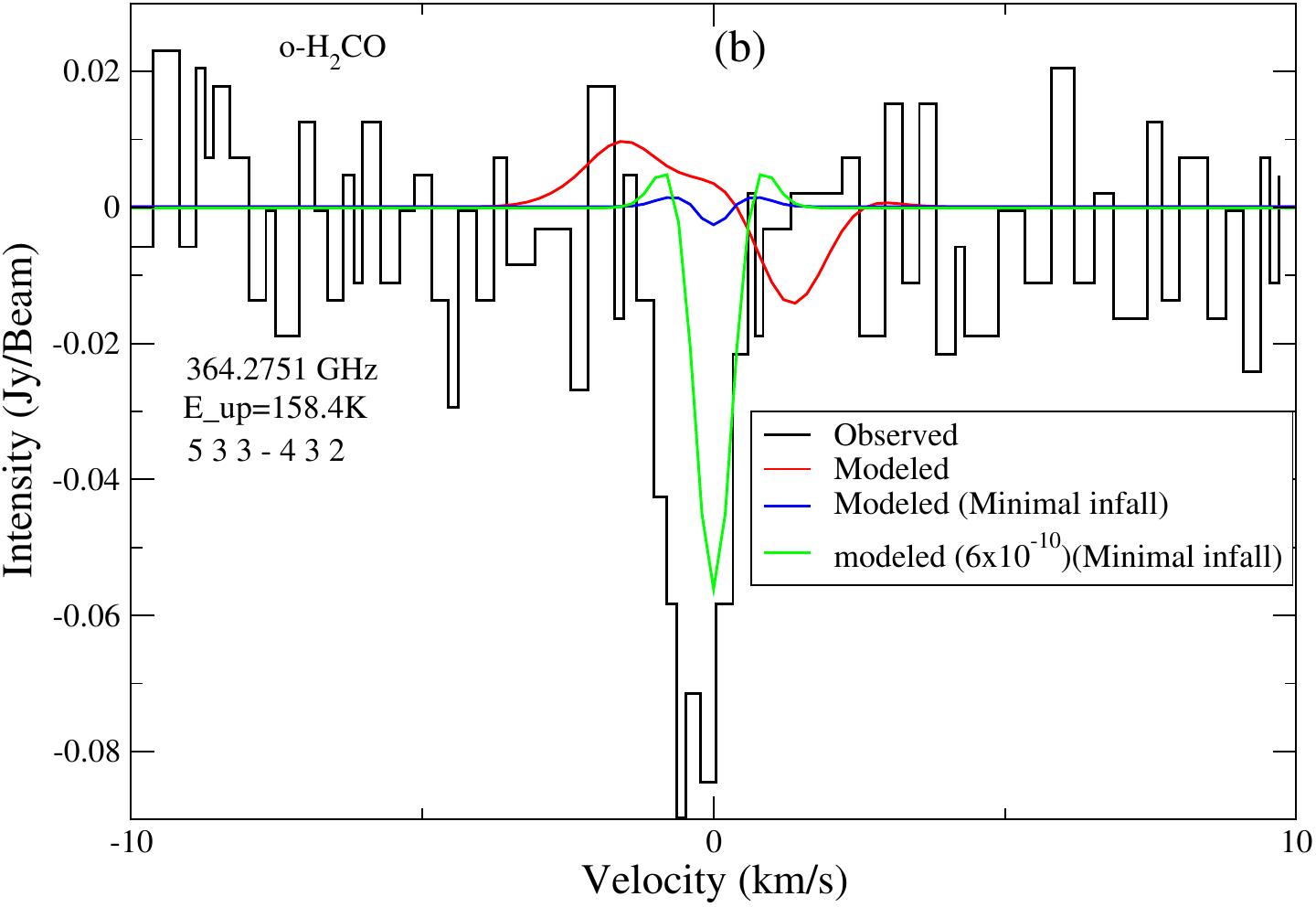} \end{minipage} 
\begin{minipage}{0.32\textwidth}
\includegraphics[width=\textwidth]{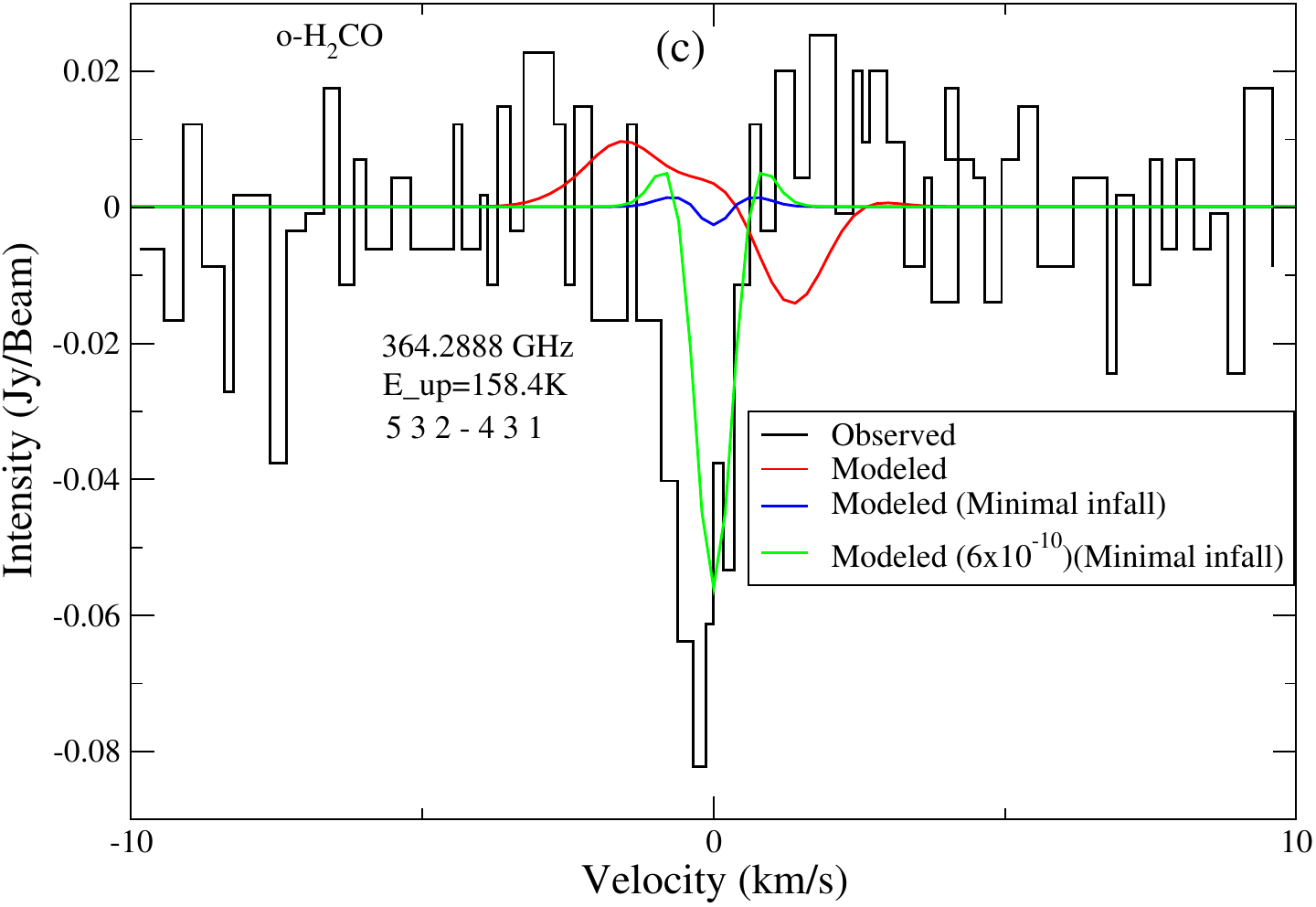} \end{minipage} 
\begin{minipage}{0.32\textwidth}
\includegraphics[width=\textwidth]{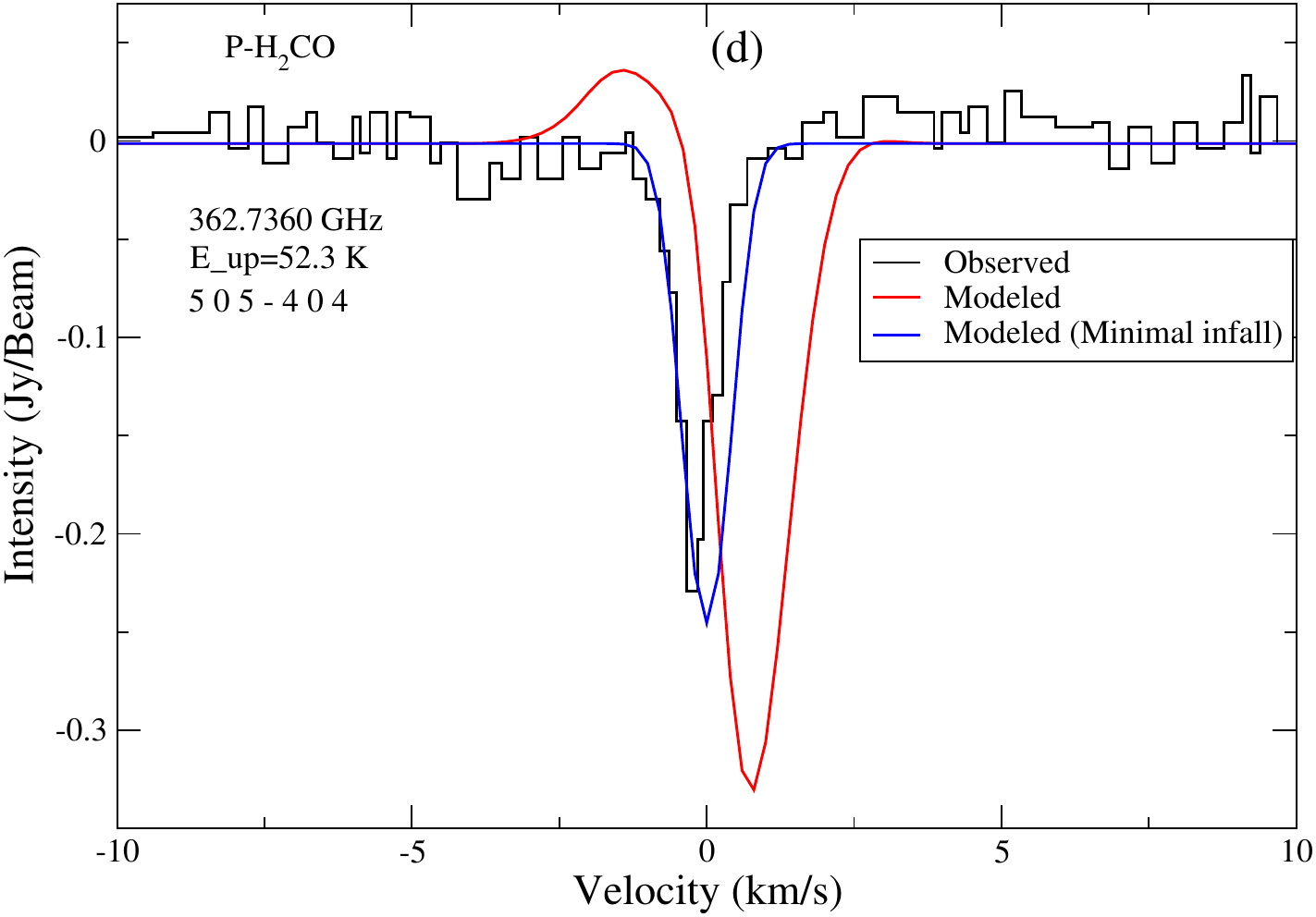} \end{minipage}
\begin{minipage}{0.32\textwidth}
\includegraphics[width=\textwidth]{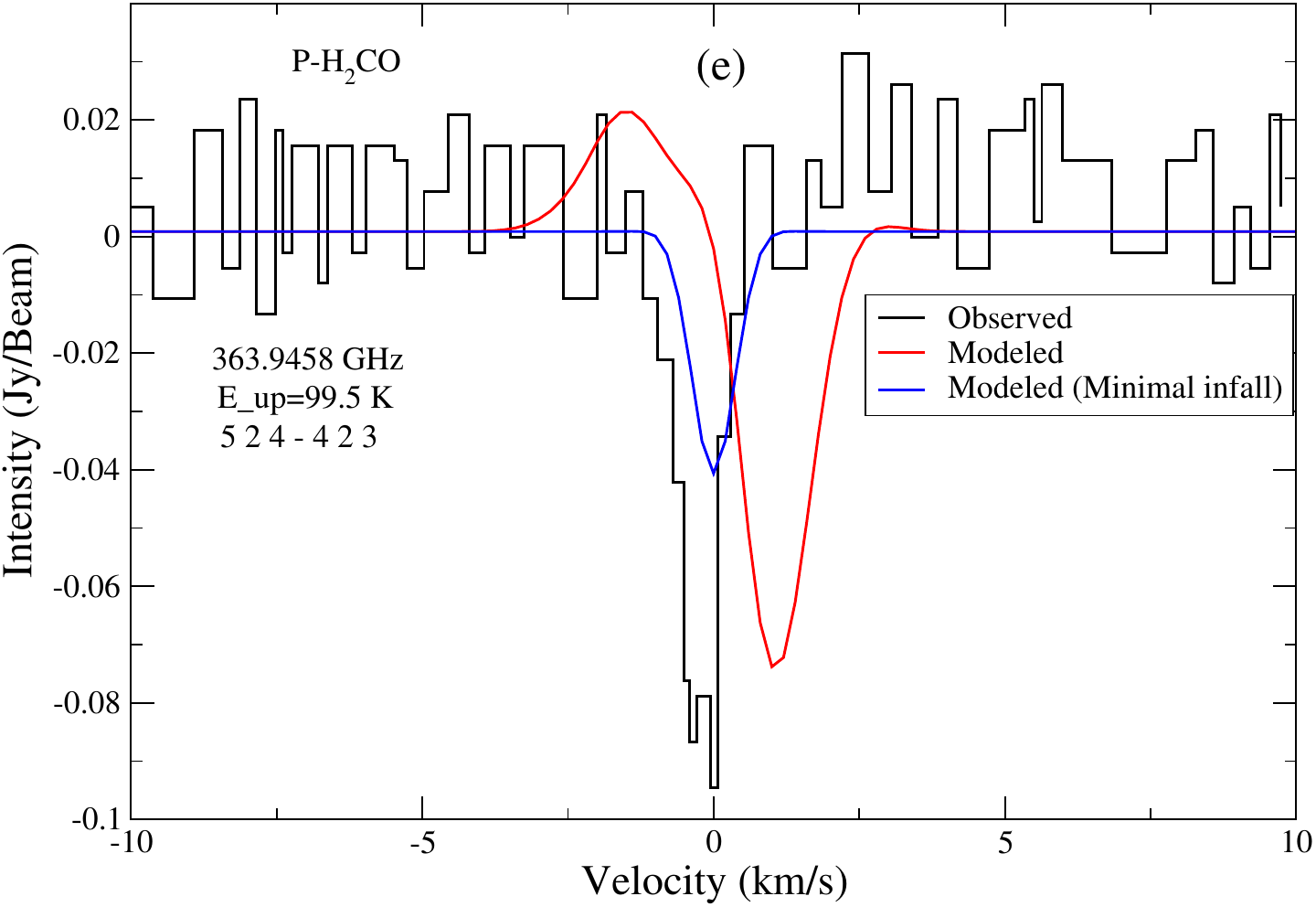} \end{minipage}
\begin{minipage}{0.32\textwidth}
\includegraphics[width=\textwidth]{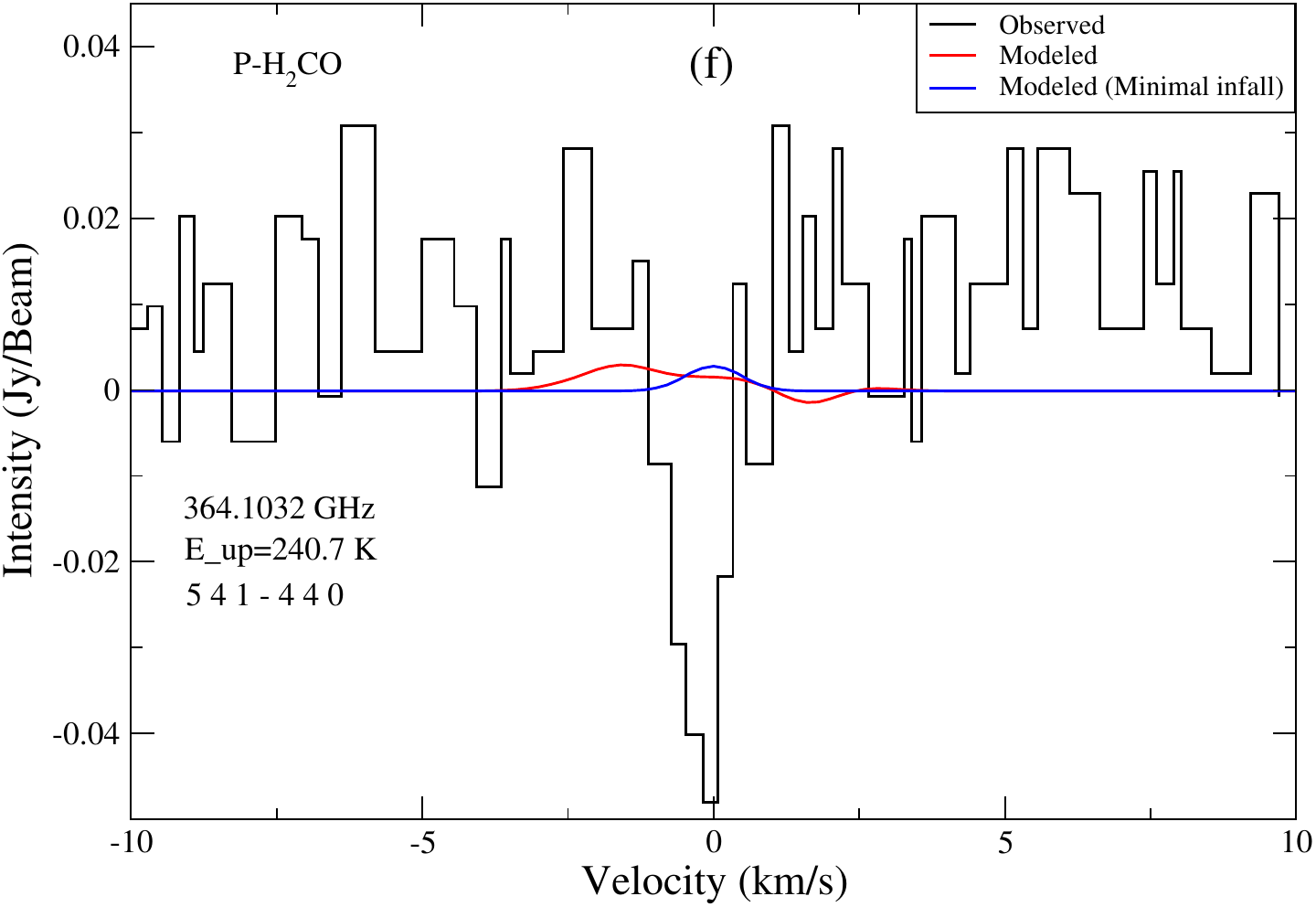} \end{minipage}
\caption{A comparison between the observed H$_2$CO lines (black) towards IRAS4A1 core, with modeled (red) line profiles for (a) o-H$_2$CO, $5_{1,5} \rightarrow 4_{1,4}$ transition (b) o-H$_2$CO, $5_{3,3} \rightarrow 4_{3,2}$ transition (c) o-H$_2$CO, $5_{3,2} \rightarrow 4_{3,1}$ (d) p-H$_2$CO, $5_{0,5} \rightarrow 4_{0,4}$ transition (e) p-H$_2$CO, $5_{2,4} \rightarrow 4_{2,3}$ transition (f) p-H$_2$CO, $5_{4,1} \rightarrow 4_{4,0}$ transition using a constant abundance $1.1\times10^{-10}$ and $6\times10^{-11}$ for o-H$_2$CO and p-H$_2$CO respectively. A FWHM value 0.90 km s$^{-1}$ is used for both o-H$_2$CO and p-H$_2$CO. The Blue lines are modeled line profiles with the same conditions; only the infall velocity considered is V$_{1000}$=0.01 (nearly static envelope). The green lines are modeled line profiles with a slightly higher abundance of H$_2$CO, $6\times10^{-10}$. \label{fig:A1_h2co}}
\end{figure*}

\subsubsection{\rm H$_2$CO}
Similar to the CH$_3$OH lines observed by \citet{sahu19} in absorption towards A1, \citet{suyu19} detected six transitions of H$_2$CO in absorption towards A1, which were also observed in emission towards A2. The upstate energy ranges from 52K to 241K for six transitions of H$_2$CO. We implemented radiative transfer modeling using RATRAN code to generate the synthetic spectra for the observed transitions of H$_2$CO. We used the same model parameters described in the Section \ref{sec:ratran}. The collisional data file for both o-H$_2$CO and p-H$_2$CO is taken from the LAMDA database (\url{https://home.strw.leidenuniv.nl/~moldata/}), originally from \citet{wies13}, with two collision partners o-H$_2$ and p-H$_2$. We used ortho and para H$_2$ as collision partners for both o-H$_2$CO and p-H$_2$CO. Initially, we used  a constant abundance of $\sim 1.1\times10^{-10}$ for o-H$_2$CO and $6\times10^{-11}$ for p-H$_2$CO and a FWHM of 0.9 km s$^{-1}$. In Figure. \ref{fig:A1_h2co} all the H$_2$CO lines observed towards A1 are shown (black) along with the spectra obtained from our radiative transfer calculation (red). The synthetic spectra appear asymmetric relative to the LSR velocity when infall is present in our model (red). However, the observed spectra (black) are symmetric and purely in absorption. When the infall velocity is set to a minimal value, V$_{1000}$=0.01 in our model, the synthetic spectra become symmetric and purely in absorption (blue), which better matches the observed spectra. Specifically, the transitions with low upper state energy of O-H$_2$CO (see Fig. \ref{fig:A1_h2co}a with $E_u = 62.5$ K) and p-H$_2$CO (see \ref{fig:A1_h2co}d with $E_u=52.3$ K) depict a better match. 
Higher energy transitions (high upper state energy transitions) typically originate from the inner regions of the envelope and require a greater abundance of H$_2$CO to explain their observed intensity. 
For o-H$_2$CO, the higher upper state energy transitions such as $5_{3,3}$ - $4_{3,2}$ and $5_{3,2}$ - $4_{3,1}$ would need to consider a higher abundance of H$_2$CO to match the spectra with the observation. 
In Figures \ref{fig:A1_h2co}b and \ref{fig:A1_h2co}c, the intensity of the model spectra, which has a higher abundance of approximately $6\times10^{-10}$ (shown in green), matches the observations fairly well. However, we could not match the two transitions having high upper state energy depicted in Figures \ref{fig:A1_h2co}e and \ref{fig:A1_h2co}f for p-H$_2$CO. In analyzing the rotational diagram of observed H$_2$CO transitions towards the A1 core, \citet{suyu19} estimated the column density of H$_2$CO to be $3\times10^{14}$ cm$^{-2}$. \citet{sahu19} calculated the H$_2$ column density towards A1 to be $1.30\times10^{26}$ cm$^{-2}$. This results in an abundance of $2.31\times10^{-12}$, which is one or two orders lower than the best-fitted abundance value derived from the RATRAN modeling of the observed H$_2$CO lines, as noted in Table \ref{table:RATRAN-best-fit}.

\begin{figure*}  
\centering
\includegraphics[width=1.0\textwidth]{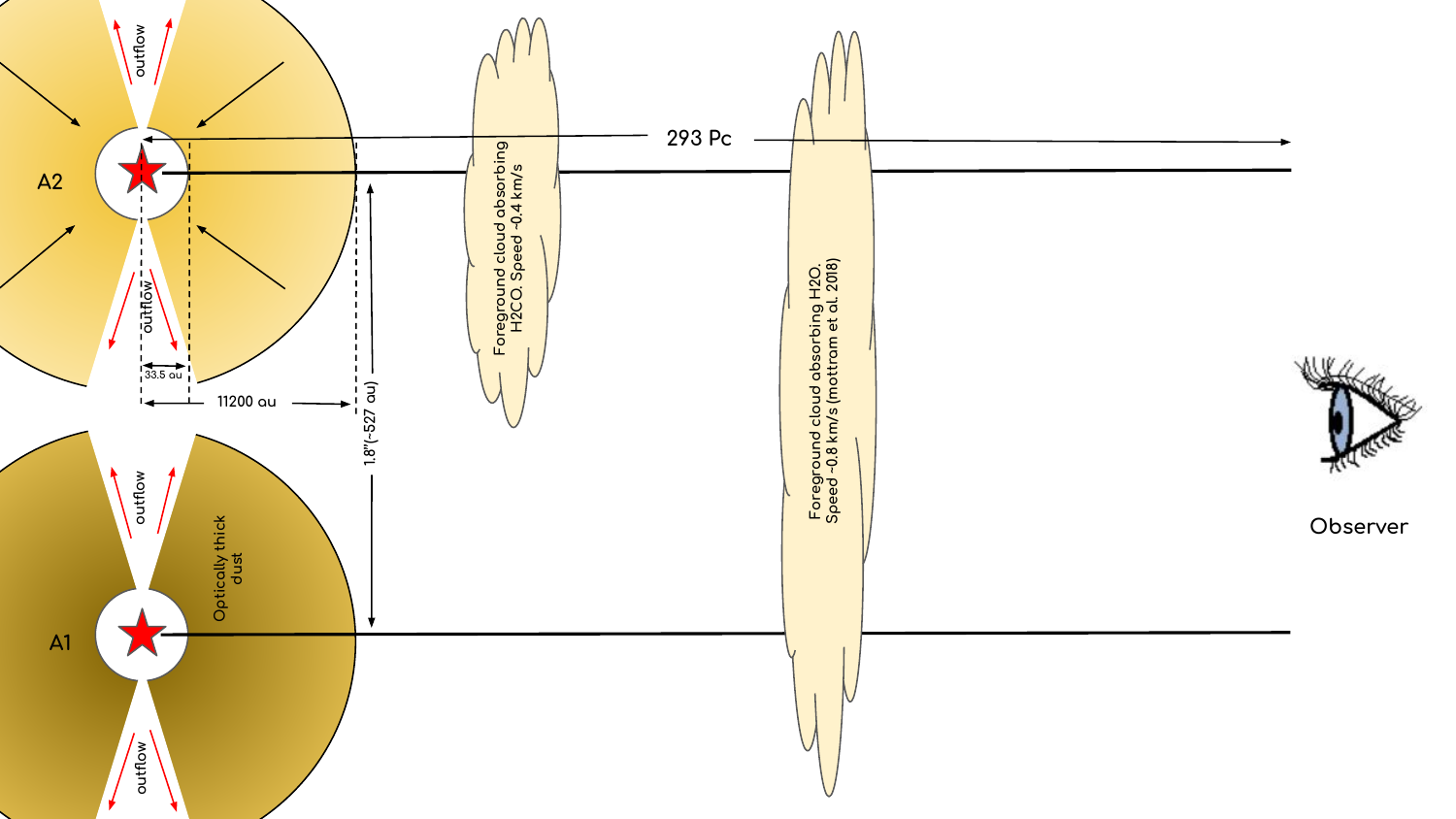}
\caption{Schematic diagram for NGC 1333 IRAS4A protobinary system. \label{fig:cartoon}}
\end{figure*}

For IRAS4A2, the inclusion of an infall velocity component is essential to reproduce the observed spectral asymmetries, suggesting the presence of inward motions in the protostellar envelope. Conversely, in the case of IRAS4A1, models incorporating same infall velocity as considered in A2, fail to reproduce the observed profiles, while a nearly static envelope (V$_{1000}$=0.01) assumption provides a better agreement. This difference may arise from observational limitations, in particular the effects of beam dilution at the employed angular resolution, which could suppress or smear out weaker kinematic signatures such as infall motions. \citet{sant15} observed infall in both the A1 and A2 components, where A1 and A2 are at different evolutionary stages. A1 is much younger and is in a very early stage of star formation, prior to the hot corino phase, while A2 is more evolved. This difference likely results in less prominent infall in A1 compared to A2.

Through our radiative transfer modeling and fitting observed line profiles, we derived the kinematical scenario of the IRAS4A system. The model's pictorial representation of the IRAS4A system is depicted in Figure \ref{fig:cartoon}. Although we did not include outflow in our model for IRAS4A1, it has already been detected by \citet{suyu19,guer24} and is therefore represented in the diagram.

\subsection{CH$_3$OH lines in centimeter wavelength}

\begin{table}
\centering
{\scriptsize
 \caption{Detected CH$_3$OH lines towards IRAS4A from observation in centimeter wavelength. \label{table:cm_observation}.}
\begin{tabular}{|l|l|l|l|}
  \hline
  \hline
  Frequency (GHz)&Qn NO.&E$_{up}$ (K)& log A$_{ij}$\\
  \hline
  \hline
24.9287(e) &3(2,1) - 3(1,2)&36.17&-7.14\\
24.9334(e)&4(2,2) - 4(1,3)&45.45&-7.10\\
24.9344(e)&2(2,0) - 2(1,1)&29.20&-7.24\\
24.9591(e)&5(2,3) - 5(1,4)&57.06&-7.08\\
25.0181(e)&6(2,4) - 6(1,5)&71.00&-7.06\\
25.1249(e)&7(2,5) - 7(1,6)&87.25&-7.04\\
25.2944(e)&8(2,6) - 8(1,7)&105.83&-7.03\\
25.5414(e)&9(2,7) - 9(1,8)&126.74&-7.01\\
25.8783(e)&10(2,8) - 10(1,9)&149.97&-6.98\\
26.3131(e)&11(2,9) - 11(1,10)&175.52&-6.95\\
\hline
\hline
\end{tabular}}\\
\end{table}
\begin{figure*}
\centering
\includegraphics[height=11cm]{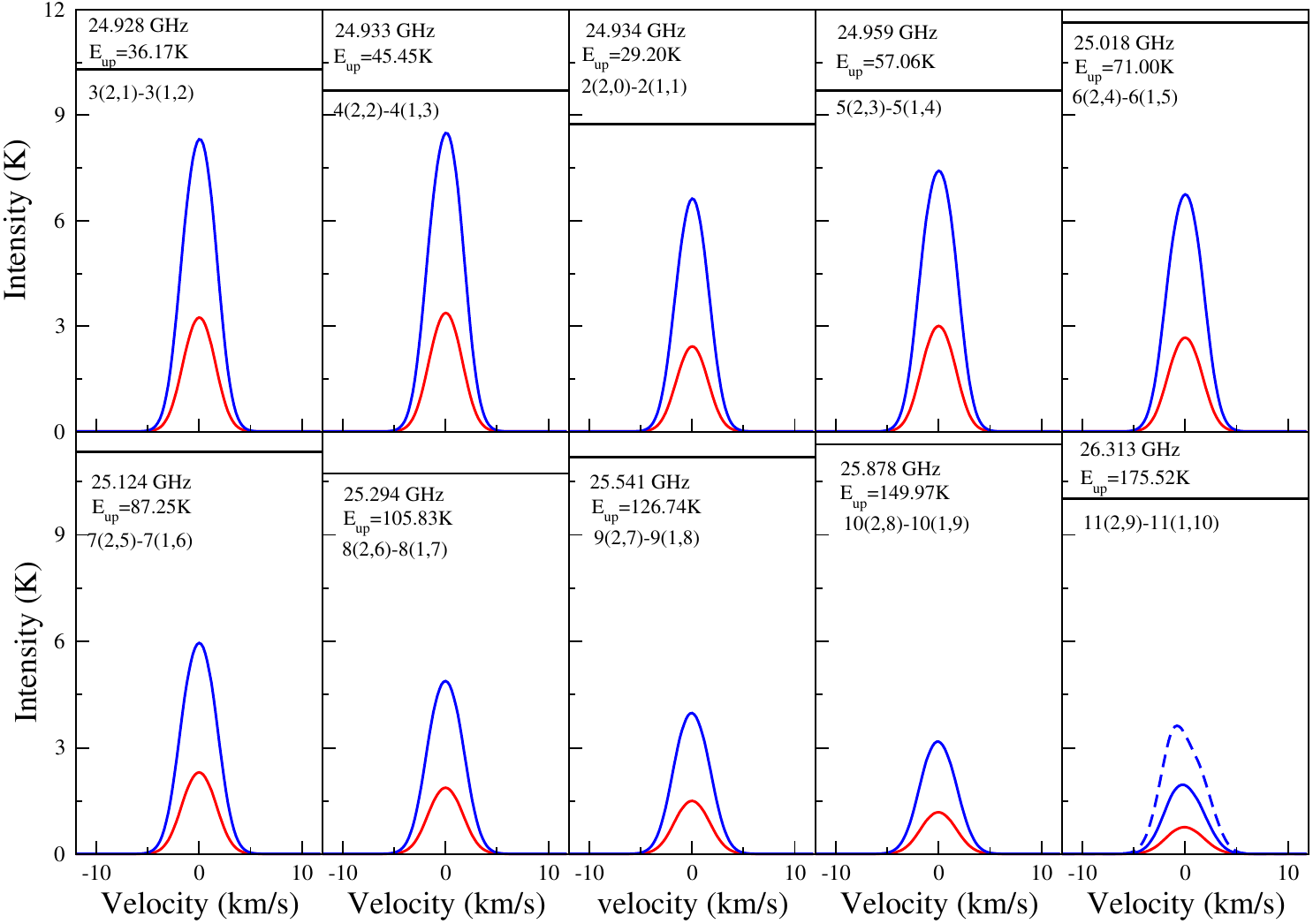}
  \caption{Synthetic spectra obtained from RATRAN modeling for CH$_{3}$OH lines in centimeter wavelength towards IRAS4A2. The red lines are using CH$_3$OH abundance $8\times10^{-9}$ from Table \ref{table:RATRAN-best-fit} in our model, where blue lines are using  CH$_{3}$OH abundance $2.5\times10^{-8}$ in our model. The horizontal black line is the observed intensity from \citet{des20a}. \label{fig:a2_desi} }
\end{figure*}

\begin{figure*}
\centering
\includegraphics[height=11cm]{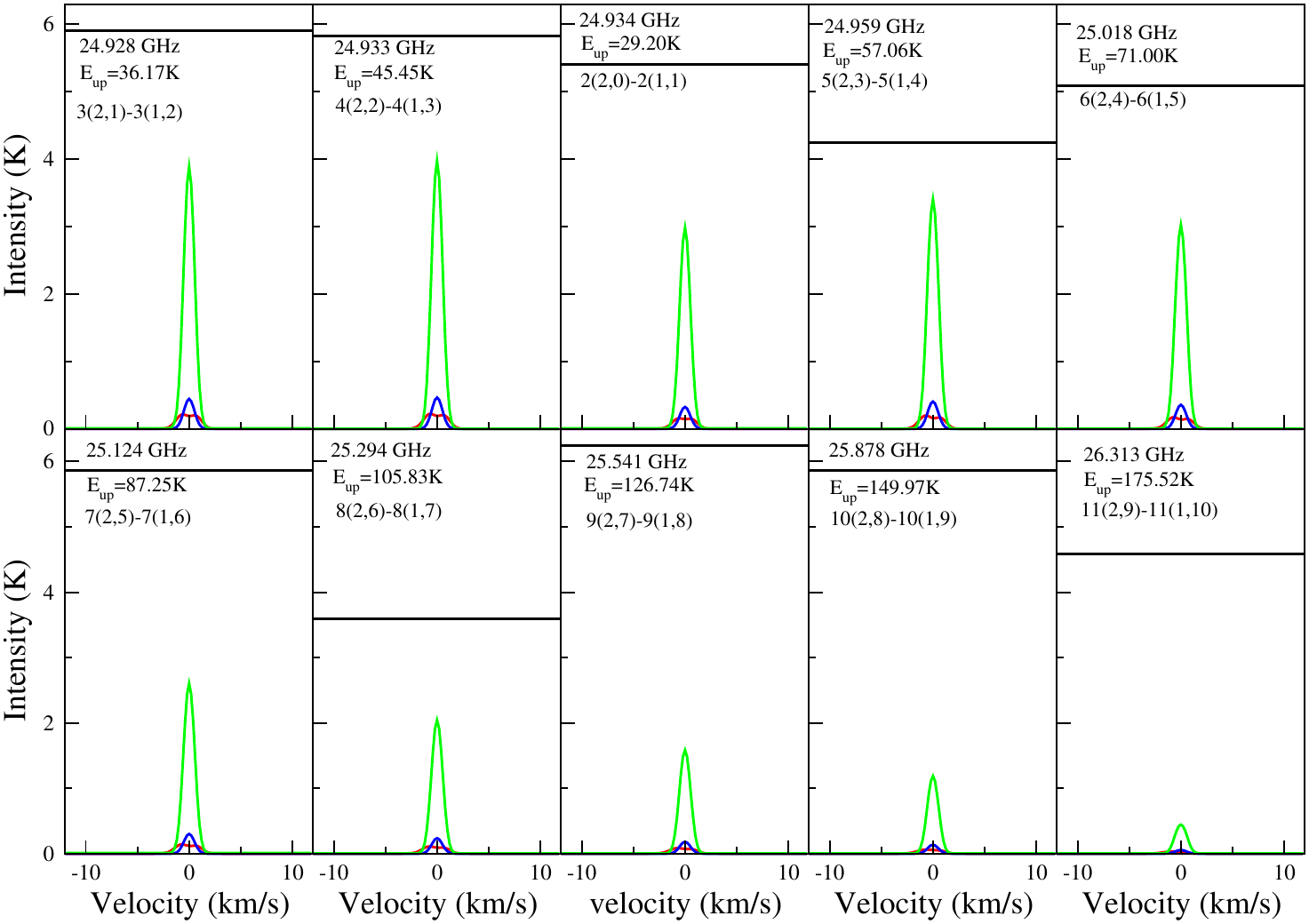}
  \caption{Synthetic spectra obtained from RATRAN modeling for CH$_{3}$OH lines in centimeter wavelength towards IRAS4A1. The red lines are using CH$_3$OH abundance $4\times10^{-10}$ from Table \ref{table:RATRAN-best-fit} in our model, where blue lines are using the same abundance but considering 0 infall velocity of the envelope. The green line is the synthetic spectra with a higher CH$_{3}$OH abundance of $4\times10^{-9}$ and no infall present in the envelope. The horizontal black line is the observed intensity from \citet{des20a}. \label{fig:a1_desi_new}}
\end{figure*}

The IRAS4A system was subsequently observed by \citet{des20a} in the centimeter wavelength region using K-band observations from the Very Large Array (VLA) telescope. This observation at 1.3 cm is significant because the dust opacity at this wavelength is negligible. Previous millimeter observations indicated that the 4A1 component lacks iCOMs lines, while \citet{sahu19,suyu19} reported the presence of these lines in absorption towards 4A1.  In contrast, these lines were observed in emission towards 4A2. The high dust opacity in the 4A1 core likely accounts for this discrepancy. We applied our radiative transfer model to generate synthetic spectra for the CH$_3$OH lines in the centimeter regime as observed by \citet{des20a} (see Table \ref{table:cm_observation}).
Figure \ref{fig:a2_desi} shows the synthetic spectra from RATRAN for the CH$_3$OH centimeter lines towards 4A2, depicted in red. These lines are in emission, which is consistent with expectations and corresponds to a constant abundance of $8\times10^{-9}$ (as indicated in Table \ref{table:RATRAN-best-fit}). The horizontal black line is the observed value by \citet{des20a} for all the transitions listed in table \ref{table:cm_observation}. The observed peak intensity, indicated by the black horizontal line, is derived from the integrated intensity reported in \citet{des20a} using the formula:
Integrated intensity = peak intensity$\times$ FWHM $\times$ 1.064. Blue spectra are the modeled line profiles with a higher abundance of CH$_3$OH, $2.5\times10^{-8}$. The modeled intensity for transitions with upper-state energy below 60 K (see the first four panels of Fig. \ref{fig:a2_desi}) aligns very well. However, it is underproduced for transitions with upper-state energy above 60 K. More interestingly, in Figure \ref{fig:a1_desi_new}, the synthetic spectra for centimeter CH$_3$OH lines are obtained using the same radiative transfer model towards 4A1, which is in emission. The red spectra represent scenarios with infall, while the blue spectra indicate conditions without infall in our model, both assuming a constant CH$_3$OH abundance of $4\times10^{-10}$. The black horizontal line is the observed intensity by \citet{des20a}. 
The green curve represents the modeled spectra with a higher abundance value of $4 \times 10^{-9}$. This value is similar to the observed value; however, the spectra in the last three panels are underproduced because these lines have higher upper-state energy, i.e., from 126-176 K,  compared to other transitions with low upper-state energy, i.e., from 36-105 K, likely originating from a more inner part of the core. 

As observed in the centimeter observations, the lines towards 4A1 are detected in emission, whereas the CH$_3$OH lines at millimeter wavelengths are observed in absorption towards 4A1. Our radiative transfer model for IRAS4A corroborates these findings, demonstrating the model's validity for both millimeter and centimeter observations. 

This helps clarify that high dust opacity is the reason for obscuring the detection of iCOMs towards 4A1. The opacity effect is significantly lower at centimeter wavelengths following 
$\mathbf{\kappa_\nu \propto \nu^{\beta}}$, where $\kappa_\nu$ is the dust opacity at frequency $\nu$ and $\beta$ is 1.0 (Table \ref{table:RATRAN-best-fit}) from our model.


\subsection{Maser lines in cm wavelength}

\begin{figure*}
\centering
\includegraphics[height=11cm]{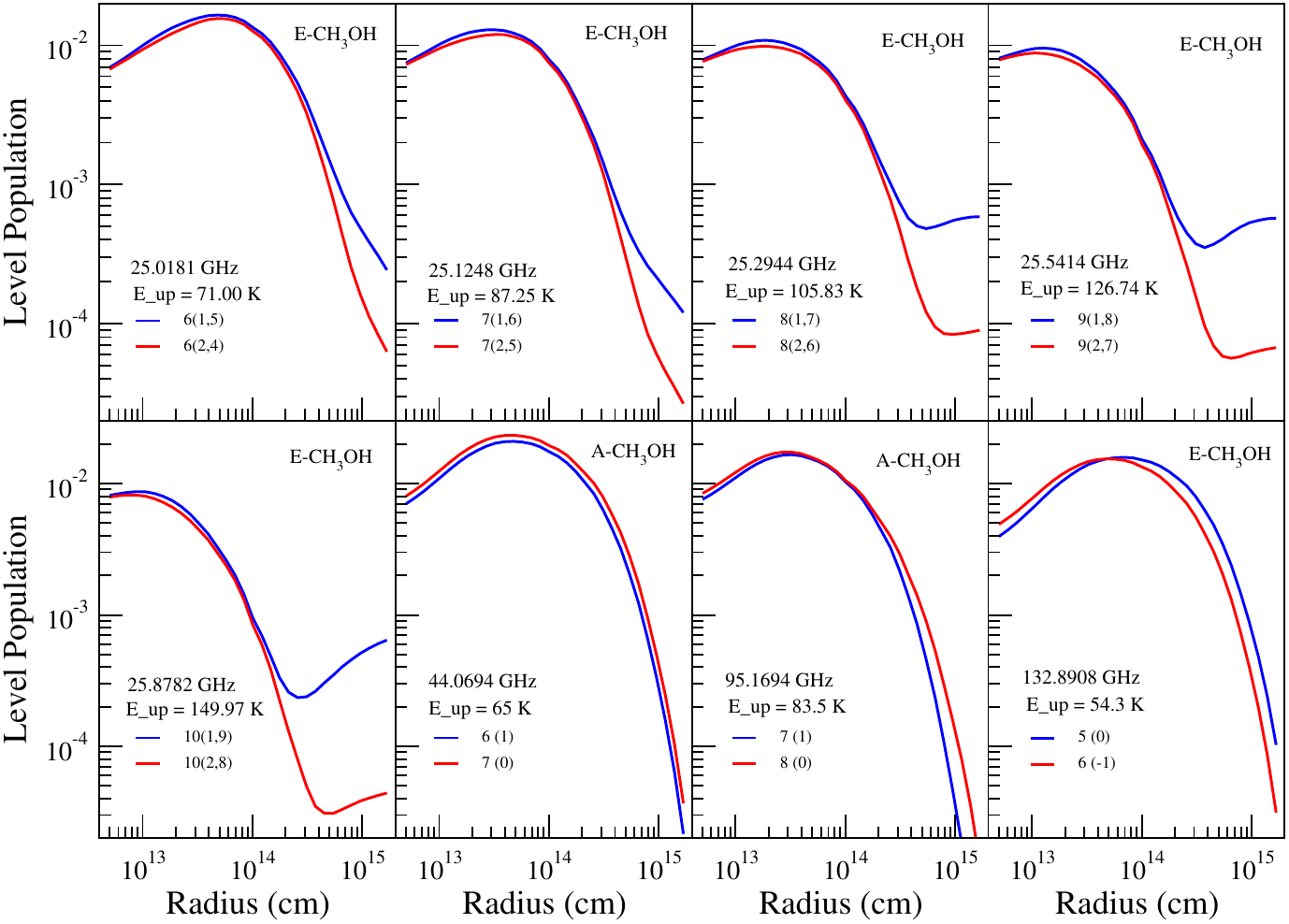}
  \caption{The blue line indicates the level population for lower levels, and the red lines indicate the population for upper levels of CH$_3$OH maser lines in centimeter wavelength towards IRAS4A2 from the RATRAN model. \label{fig:maser}}
\end{figure*}

Outflows and jets are the most common phenomena during star formation. This energetic process significantly impacts both star formation and the accretion of matter in the circumstellar disk. The bright 25 GHz series of methanol masers, which are produced in the energetic regions around massive stars, serve as natural indicators of shocked gas surrounding newly forming stars. Outflows towards the IRAS4A system have been previously detected by \citet{des20b}. This source is a promising target for studying maser lines, which can help us better understand the outflow and shock activities in the region. \citet{des20a} conducted observations of a series of 25 GHz lines towards IRAS4A using the k-band receivers of the VLA; however, these lines were not detected with maser actions. The lack of detection is mainly attributed to the physical parameters of the two regions, including gas density, temperature, and methanol column density, which are not suitable enough for population inversion. In practice, since the energy levels are populated according to local thermal equilibrium (LTE) and the lines are optically thick, there is no inversion present. In Figure \ref{fig:maser} first five panels show the population for the lower (blue line) and upper levels (red line) for the five transitions around 25 GHz listed in Table \ref{table:cm_observation}. These level populations are obtained from our radiative transfer model towards IRAS4A2. From these figures, we can see the population for lower levels is greater than the upper level throughout the cloud, which again suggests that for these lines, there is no population inversion and this is the reason for not observing maser action among these lines. \citet{lyo14} observed 44 GHz class I CH$_{3}$OH maser line towards IRAS4A2 using the KVN 21 m telescopes. They also detected 95.16946 GHz (8$_{0}$ – 7$_{1}$ A$^{+}$), and 132.89080 GHz (6$_{−1}$ – 5$_{0}$ E) maser transitions of CH$_{3}$OH towards IRAS4A2. We obtained the level populations for three transitions using our RATRAN modeling, which we presented in the last three panels of Figure \ref{fig:maser}. It appears that, in most regions, the lower level (represented by the blue line) is less populated compared to the upper level for the 44 GHz and 95 GHz lines. This difference in population is a criterion for observing these lines in maser action. For the 132 GHz line, only in the inner region, the lower level (blue) is less populated, whereas in the outer region, no population inversion occurs.

\section{Conclusions}\label{sec:concl}
This paper presents a simple 1D spherical radiative transfer model aimed at explaining several observed line profiles in the low-mass protobinary system IRAS4A, which consists of two components: A1 and A2. Previous ALMA observations at millimeter wavelengths have revealed striking spectral differences between these two components. We have conducted a multi-wavelength study to validate our model across both millimeter and centimeter wavelength regimes. The results can be summarized as follows-:

\begin{itemize}

\item {
Our radiative transfer models enable us to extract various physical properties, such as infall velocity, mass inflow rate, and full width at half maximum (FWHM). The spectral differences between regions A1 and A2 are accurately represented in our model at millimeter wavelengths. To match the model spectra with the observed data, we incorporated a temperature gradient by representing temperature stratified layers as a spherical shell in our model. This approach allowed us to consider one possible scenario described by \citet{sahu19} regarding A1. However, we could not account for an alternative scenario involving a hot, optically thick disk due to certain limitations within our model. Ultimately, our model successfully represents the observed absorption lines toward A1 with these considerations in mind.}

\item {Our analysis showed that dust opacity is higher toward A1 compared to A2. The differences in the abundance of iCOMs observed between A1 and A2 in the millimeter (mm) regime primarily result from the high dust opacity surrounding A1. This high opacity obscures molecular lines, making it difficult to accurately measure their abundance, which causes them to appear underproduced. In contrast, the lower dust opacity in the centimeter (cm) regime allows for unobscured molecular emissions, leading to higher measured abundances. We extended our model to the cm regime and successfully matched the observed lines of CH$_3$OH observed in emission towards A1 reported by de \citet{des20a}. Our radiative transfer model effectively explains the line profiles observed in both the mm and cm regimes. Using this model, we determined that the abundance of CH$_3$OH toward A1 is approximately 10$^{-10}$ in the mm regime, while a higher best-fit abundance of 10$^{-9}$ is found at centimeter wavelengths.}

\item { We obtained a best-fit infall velocity at a distance of 1000 au of 0.55 km s$^{-1}$, which is less than the value obtained by \citet{mot13}, which was 1.1 km s$^{-1}$. A possible reason for this discrepancy is the difference in the maximum recoverable scale in the observations. The impact of the MRS on our results cannot be quantitatively incorporated in this work, as our analysis is based on a simplified toy model. The upper state energy of the observed H$_2$O transition using HIFI \cite{mot13} ranges from $50$K to $250$K, while CH$_3$OH and H$_2$CO lines observed with ALMA range from 16-241K and 52-241K, respectively. Different molecular tracers affect critical density and excitation conditions. Methanol often traces larger scales ($\sim$0.3–1.0 pc), leading to potentially lower infall velocities than water, which samples smaller, denser regions. Methanol lines in the mm/sub-mm range are generally more reliable for infall measurements, while water lines may reflect both infall and outflow processes.}

\item { To explain the observed line profiles toward IRAS4A2, we take into account the presence of an infall velocity in the source. In contrast, for IRAS4A1, incorporating the same infall velocity as A2 in our model does not correlate with the observed profiles. However, when we set the infall velocity to a minimal value in our model, we can effectively explain the observed line profiles for IRAS4A1. A possible explanation could be the angular resolution of the observation. The molecular emission from the region might be getting diluted due to the larger beam size.}

\item {Our model shows that the CH$_3$OH maser lines at 44 GHz and 95 GHz, which were previously observed, exhibit a higher population in the upper levels of the transitions. Our model demonstrated this population inversion for the observed maser lines.}

\end{itemize}

\section{acknowledgment}
This paper makes use of the following ALMA  data: ADS/JAO.ALMA $2015.1.00147.S$. ALMA is a partnership of ESO (representing its member states),
NSF (USA)and NINS (Japan), together with NRC (Canada), MOST and  ASIAA  (Taiwan), and  KASI  (Republic of Korea), in cooperation with the Republic of Chile, operate the Joint ALMA Observatory. ESO, AUI/NRAO, and NAOJ also operate it. This work was carried out at the Physical Research Laboratory (PRL), Ahmedabad, and the Indian Institute of Astrophysics (IIA), Bengaluru. B.B. thanks both institutes for providing research facilities and support. D.S. acknowledges the Ramanujan Fellowship grant and PRL, Ahmedabad, for continued research support.

\clearpage

\section{Bibliography}



\bibliographystyle{cas-model2-names}

\bibliography{references}

@ARTICLE{hoge00,
       author = {{Hogerheijde}, M.~R. and {van der Tak}, F.~F.~S.},
        title = "{An accelerated Monte Carlo method to solve two-dimensional radiative transfer and molecular excitation. With applications to axisymmetric models of star formation}",
      journal = {aap},
         year = "2000",
        month = "Oct",
       volume = {362},
        pages = {697-710},
archivePrefix = {arXiv},
       eprint = {astro-ph/0008169},
 primaryClass = {astro-ph},
       adsurl = {https://ui.adsabs.harvard.edu/abs/2000A\&A...362..697H}
}

@ARTICLE{rabl10,
       author = {{Rabli}, Djamal and {Flower}, D.~R.},
        title = "{The rotational excitation of methanol by molecular hydrogen}",
      journal = {mnras},
         year = "2010",
        month = "Jul",
       volume = {406},
       number = {1},
        pages = {95-101},
          doi = {10.1111/j.1365-2966.2010.16671.x},
       adsurl = {https://ui.adsabs.harvard.edu/abs/2010MNRAS.406...95R}
}

@ARTICLE{vand07,
       author = {{van der Tak}, F.~F.~S. and {Black}, J.~H. and {Sch{\"o}ier}, F.~L. and
         {Jansen}, D.~J. and {van Dishoeck}, E.~F.},
        title = "{A computer program for fast non-LTE analysis of interstellar line spectra. With diagnostic plots to interpret observed line intensity ratios}",
      journal = {aap},
         year = "2007",
        month = "Jun",
       volume = {468},
       number = {2},
        pages = {627-635},
          doi = {10.1051/0004-6361:20066820},
archivePrefix = {arXiv},
       eprint = {0704.0155},
 primaryClass = {astro-ph},
       adsurl = {https://ui.adsabs.harvard.edu/abs/2007A\&A...468..627V}
}

@ARTICLE{belt18,
       author = {{Beltr{\'a}n}, M.~T. and {Cesaroni}, R. and {Rivilla}, V.~M. and
         {S{\'a}nchez-Monge}, {\'A}. and {Moscadelli}, L. and {Ahmadi}, A. and
         {Allen}, V. and {Beuther}, H. and {Etoka}, S. and {Galli}, D. and
         {Galv{\'a}n-Madrid}, R. and {Goddi}, C. and {Johnston}, K.~G. and
         {Klaassen}, P.~D. and {K{\"o}lligan}, A. and {Kuiper}, R. and
         {Kumar}, M.~S.~N. and {Maud}, L.~T. and {Mottram}, J.~C. and
         {Peters}, T. and {Schilke}, P. and {Testi}, L. and {van der Tak}, F. and
         {Walmsley}, C.~M.},
        title = "{Accelerating infall and rotational spin-up in the hot molecular core G31.41+0.31}",
      journal = {aap},
         year = 2018,
        month = jul,
       volume = {615},
          eid = {A141},
        pages = {A141},
          doi = {10.1051/0004-6361/201832811},
archivePrefix = {arXiv},
       eprint = {1803.05300},
 primaryClass = {astro-ph.SR},
       adsurl = {https://ui.adsabs.harvard.edu/abs/2018A\&A...615A.141B}
}

@ARTICLE{mot13,
       author = {{Mottram}, J.~C. and {van Dishoeck}, E.~F. and {Schmalzl}, M. and
         {Kristensen}, L.~E. and {Visser}, R. and {Hogerheijde}, M.~R. and
         {Bruderer}, S.},
        title = "{Waterfalls around protostars. Infall motions towards Class 0/I envelopes as probed by water}",
      journal = {aap},
         year = "2013",
        month = "Oct",
       volume = {558},
          eid = {A126},
        pages = {A126},
          doi = {10.1051/0004-6361/201321828},
archivePrefix = {arXiv},
       eprint = {1308.5119},
 primaryClass = {astro-ph.GA},
       adsurl = {https://ui.adsabs.harvard.edu/abs/2013A\&A...558A.126M}
}

@ARTICLE{oss94,
       author = {{Ossenkopf}, V. and {Henning}, Th.},
        title = "{Dust opacities for protostellar cores.}",
      journal = {aap},
         year = "1994",
        month = "Nov",
       volume = {291},
        pages = {943-959},
       adsurl = {https://ui.adsabs.harvard.edu/abs/1994A\&A...291..943O}
}

@ARTICLE{van07,
       author = {{van der Tak}, F.~F.~S. and {Black}, J.~H. and {Sch{\"o}ier}, F.~L. and
         {Jansen}, D.~J. and {van Dishoeck}, E.~F.},
        title = "{A computer program for fast non-LTE analysis of interstellar line spectra. With diagnostic plots to interpret observed line intensity ratios}",
      journal = {aap},
         year = "2007",
        month = "Jun",
       volume = {468},
       number = {2},
        pages = {627-635},
          doi = {10.1051/0004-6361:20066820},
archivePrefix = {arXiv},
       eprint = {0704.0155},
 primaryClass = {astro-ph},
       adsurl = {https://ui.adsabs.harvard.edu/abs/2007A\&A...468..627V}
}

@INPROCEEDINGS{vast15,
       author = {{Vastel}, C. and {Bottinelli}, S. and {Caux}, E. and {Glorian}, J. -M. and {Boiziot}, M.},
        title = "{CASSIS: a tool to visualize and analyse instrumental and synthetic spectra.}",
    booktitle = {SF2A-2015: Proceedings of the Annual meeting of the French Society of Astronomy and Astrophysics},
         year = 2015,
        month = dec,
        pages = {313-316},
       adsurl = {https://ui.adsabs.harvard.edu/abs/2015sf2a.conf..313V}
}

@ARTICLE{difr01,
       author = {{Di Francesco}, James and {Myers}, Philip C. and {Wilner}, David J. and {Ohashi}, Nagayoshi and {Mardones}, Diego},
        title = "{Infall, Outflow, Rotation, and Turbulent Motions of Dense Gas within NGC 1333 IRAS 4}",
      journal = {apj},
         year = 2001,
        month = dec,
       volume = {562},
       number = {2},
        pages = {770-789},
          doi = {10.1086/323854},
archivePrefix = {arXiv},
       eprint = {astro-ph/0108022},
 primaryClass = {astro-ph},
       adsurl = {https://ui.adsabs.harvard.edu/abs/2001ApJ...562..770D}
}

@ARTICLE{kris12,
       author = {{Kristensen}, L.~E. and {van Dishoeck}, E.~F. and {Bergin}, E.~A. and {Visser}, R. and {Y{\i}ld{\i}z}, U.~A. and {San Jose-Garcia}, I. and {J{\o}rgensen}, J.~K. and {Herczeg}, G.~J. and {Johnstone}, D. and {Wampfler}, S.~F. and {Benz}, A.~O. and {Bruderer}, S. and {Cabrit}, S. and {Caselli}, P. and {Doty}, S.~D. and {Harsono}, D. and {Herpin}, F. and {Hogerheijde}, M.~R. and {Karska}, A. and {van Kempen}, T.~A. and {Liseau}, R. and {Nisini}, B. and {Tafalla}, M. and {van der Tak}, F. and {Wyrowski}, F.},
        title = "{Water in star-forming regions with Herschel (WISH). II. Evolution of 557 GHz {}1$_{10}$-{}1$_{01}$ emission in low-mass protostars}",
      journal = {aap},
         year = 2012,
        month = jun,
       volume = {542},
          eid = {A8},
        pages = {A8},
          doi = {10.1051/0004-6361/201118146},
archivePrefix = {arXiv},
       eprint = {1204.0009},
 primaryClass = {astro-ph.SR},
       adsurl = {https://ui.adsabs.harvard.edu/abs/2012A&A...542A...8K}
}

@article{gora21,
	doi = {10.3847/1538-4357/abc9c4},
	url = {https://doi.org/10.3847/1538-4357/abc9c4},
	year = 2021,
	month = {feb},
	publisher = {American Astronomical Society},
	volume = {907},
	number = {2},
	pages = {108},
	author = {Prasanta Gorai and Ankan Das and Takashi Shimonishi and Dipen Sahu and Suman Kumar Mondal and Bratati Bhat and Sandip K. Chakrabarti},
	title = {Identification of Methyl Isocyanate and Other Complex Organic Molecules in a Hot Molecular Core, G31.41+0.31},
	journal = {The Astrophysical Journal}
}

@INPROCEEDINGS{cecc04,
       author = {{Ceccarelli}, C.},
        title = "{The Hot Corinos of Solar Type Protostars}",
    booktitle = {Star Formation in the Interstellar Medium: In Honor of David Hollenbach},
         year = 2004,
       editor = {{Johnstone}, D. and {Adams}, F.~C. and {Lin}, D.~N.~C. and {Neufeeld}, D.~A. and {Ostriker}, E.~C.},
       series = {Astronomical Society of the Pacific Conference Series},
       volume = {323},
        month = dec,
        pages = {195},
       adsurl = {https://ui.adsabs.harvard.edu/abs/2004ASPC..323..195C}
      }

@ARTICLE{sant15,
       author = {{Santangelo}, G. and {Codella}, C. and {Cabrit}, S. and {Maury}, A.~J. and {Gueth}, F. and {Maret}, S. and {Lefloch}, B. and {Belloche}, A. and {Andr{\'e}}, Ph. and {Hennebelle}, P. and {Anderl}, S. and {Podio}, L. and {Testi}, L.},
        title = "{Jet multiplicity in the proto-binary system NGC 1333-IRAS4A. The detailed CALYPSO IRAM-PdBI view}",
      journal = {aap},
         year = 2015,
        month = dec,
       volume = {584},
          eid = {A126},
        pages = {A126},
          doi = {10.1051/0004-6361/201526323},
archivePrefix = {arXiv},
       eprint = {1511.01428},
 primaryClass = {astro-ph.GA},
       adsurl = {https://ui.adsabs.harvard.edu/abs/2015A&A...584A.126S}
}

@INPROCEEDINGS{belb18,
       author = {{Beltr{\'a}n}, M.~T. and {Rivilla}, V.~M.},
        title = "{Complex Organic Molecules in Hot Molecular Cores/Corinos: Physics and Chemistry}",
    booktitle = {Science with a Next Generation Very Large Array},
         year = 2018,
       editor = {{Murphy}, Eric},
       series = {Astronomical Society of the Pacific Conference Series},
       volume = {517},
        month = dec,
        pages = {249},
archivePrefix = {arXiv},
       eprint = {1806.08137},
 primaryClass = {astro-ph.SR},
       adsurl = {https://ui.adsabs.harvard.edu/abs/2018ASPC..517..249B}
}

@ARTICLE{bell20,
       author = {{Belloche}, A. and {Maury}, A.~J. and {Maret}, S. and {Anderl}, S. and {Bacmann}, A. and {Andr{\'e}}, Ph. and {Bontemps}, S. and {Cabrit}, S. and {Codella}, C. and {Gaudel}, M. and {Gueth}, F. and {Lef{\`e}vre}, C. and {Lefloch}, B. and {Podio}, L. and {Testi}, L.},
        title = "{Questioning the spatial origin of complex organic molecules in young protostars with the CALYPSO survey}",
      journal = {aap},
         year = 2020,
        month = mar,
       volume = {635},
          eid = {A198},
        pages = {A198},
          doi = {10.1051/0004-6361/201937352},
archivePrefix = {arXiv},
       eprint = {2002.00592},
 primaryClass = {astro-ph.GA},
       adsurl = {https://ui.adsabs.harvard.edu/abs/2020A&A...635A.198B}
}

@ARTICLE{marb18,
       author = {{Marcelino}, N. and {Gerin}, M. and {Cernicharo}, J. and {Fuente}, A. and {Wootten}, H.~A. and {Chapillon}, E. and {Pety}, J. and {Lis}, D.~C. and {Roueff}, E. and {Commer{\c{c}}on}, B. and {Ciardi}, A.},
        title = "{ALMA observations of the young protostellar system Barnard 1b: Signatures of an incipient hot corino in B1b-S}",
      journal = {aap},
         year = 2018,
        month = nov,
       volume = {620},
          eid = {A80},
        pages = {A80},
          doi = {10.1051/0004-6361/201731955},
archivePrefix = {arXiv},
       eprint = {1809.08014},
 primaryClass = {astro-ph.GA},
       adsurl = {https://ui.adsabs.harvard.edu/abs/2018A&A...620A..80M}
}

@ARTICLE{des20a,
       author = {{De Simone}, Marta and {Ceccarelli}, Cecilia and {Codella}, Claudio and {Svoboda}, Brian E. and {Chandler}, Claire and {Bouvier}, Mathilde and {Yamamoto}, Satoshi and {Sakai}, Nami and {Caselli}, Paola and {Favre}, Cecile and {Loinard}, Laurent and {Lefloch}, Bertrand and {Liu}, Hauyu Baobab and {L{\'o}pez-Sepulcre}, Ana and {Pineda}, Jaime E. and {Taquet}, Vianney and {Testi}, Leonardo},
        title = "{Hot Corinos Chemical Diversity: Myth or Reality?}",
      journal = {apjl},
         year = 2020,
        month = jun,
       volume = {896},
       number = {1},
          eid = {L3},
        pages = {L3},
          doi = {10.3847/2041-8213/ab8d41},
archivePrefix = {arXiv},
       eprint = {2006.04484},
 primaryClass = {astro-ph.SR},
       adsurl = {https://ui.adsabs.harvard.edu/abs/2020ApJ...896L...3D}
}

@ARTICLE{des20b,
       author = {{De Simone}, M. and {Codella}, C. and {Ceccarelli}, C. and {L{\'o}pez-Sepulcre}, A. and {Witzel}, A. and {Neri}, R. and {Balucani}, N. and {Caselli}, P. and {Favre}, C. and {Fontani}, F. and {Lefloch}, B. and {Ospina-Zamudio}, J. and {Pineda}, J.~E. and {Taquet}, V.},
        title = "{Seeds of Life in Space (SOLIS). X. Interstellar complex organic molecules in the NGC 1333 IRAS 4A outflows}",
      journal = {aap},
         year = 2020,
        month = aug,
       volume = {640},
          eid = {A75},
        pages = {A75},
          doi = {10.1051/0004-6361/201937004},
archivePrefix = {arXiv},
       eprint = {2006.09925},
 primaryClass = {astro-ph.SR},
       adsurl = {https://ui.adsabs.harvard.edu/abs/2020A&A...640A..75D}
}

@ARTICLE{taqu20,
       author = {{Taquet}, V. and {Codella}, C. and {De Simone}, M. and {L{\'o}pez-Sepulcre}, A. and {Pineda}, J.~E. and {Segura-Cox}, D. and {Ceccarelli}, C. and {Caselli}, P. and {Gusdorf}, A. and {Persson}, M.~V. and {Alves}, F. and {Caux}, E. and {Favre}, C. and {Fontani}, F. and {Neri}, R. and {Oya}, Y. and {Sakai}, N. and {Vastel}, C. and {Yamamoto}, S. and {Bachiller}, R. and {Balucani}, N. and {Bianchi}, E. and {Bizzocchi}, L. and {Chac{\'o}n-Tanarro}, A. and {Dulieu}, F. and {Enrique-Romero}, J. and {Feng}, S. and {Holdship}, J. and {Lefloch}, B. and {Jaber Al-Edhari}, A. and {Jim{\'e}nez-Serra}, I. and {Kahane}, C. and {Lattanzi}, V. and {Ospina-Zamudio}, J. and {Podio}, L. and {Punanova}, A. and {Rimola}, A. and {Sims}, I.~R. and {Spezzano}, S. and {Testi}, L. and {Theul{\'e}}, P. and {Ugliengo}, P. and {Vasyunin}, A.~I. and {Vazart}, F. and {Viti}, S. and {Witzel}, A.},
        title = "{Seeds of Life in Space (SOLIS). VI. Chemical evolution of sulfuretted species along the outflows driven by the low-mass protostellar binary NGC 1333-IRAS4A}",
      journal = {aap},
         year = 2020,
        month = may,
       volume = {637},
          eid = {A63},
        pages = {A63},
          doi = {10.1051/0004-6361/201937072},
archivePrefix = {arXiv},
       eprint = {2002.05480},
 primaryClass = {astro-ph.GA},
       adsurl = {https://ui.adsabs.harvard.edu/abs/2020A&A...637A..63T}
}

@ARTICLE{lope17,
       author = {{L{\'o}pez-Sepulcre}, A. and {Sakai}, N. and {Neri}, R. and {Imai}, M. and {Oya}, Y. and {Ceccarelli}, C. and {Higuchi}, A.~E. and {Aikawa}, Y. and {Bottinelli}, S. and {Caux}, E. and {Hirota}, T. and {Kahane}, C. and {Lefloch}, B. and {Vastel}, C. and {Watanabe}, Y. and {Yamamoto}, S.},
        title = "{Complex organics in IRAS 4A revisited with ALMA and PdBI: Striking contrast between two neighbouring protostellar cores}",
      journal = {aap},
         year = 2017,
        month = oct,
       volume = {606},
          eid = {A121},
        pages = {A121},
          doi = {10.1051/0004-6361/201630334},
archivePrefix = {arXiv},
       eprint = {1707.03745},
 primaryClass = {astro-ph.GA},
       adsurl = {https://ui.adsabs.harvard.edu/abs/2017A&A...606A.121L}
}

@ARTICLE{sahu19,
       author = {{Sahu}, Dipen and {Liu}, Sheng-Yuan and {Su}, Yu-Nung and {Li}, Zhi-Yun and {Lee}, Chin-Fei and {Hirano}, Naomi and {Takakuwa}, Shigehisa},
        title = "{Implications of a Hot Atmosphere/Corino from ALMA Observations toward NGC 1333 IRAS 4A1}",
      journal = {apj},
         year = 2019,
        month = feb,
       volume = {872},
       number = {2},
          eid = {196},
        pages = {196},
          doi = {10.3847/1538-4357/aaffda},
archivePrefix = {arXiv},
       eprint = {1901.05424},
 primaryClass = {astro-ph.GA},
       adsurl = {https://ui.adsabs.harvard.edu/abs/2019ApJ...872..196S}
}

@ARTICLE{suyu19,
       author = {{Su}, Yu-Nung and {Liu}, Sheng-Yuan and {Li}, Zhi-Yun and {Lee}, Chin-Fei and {Hirano}, Naomi and {Takakuwa}, Shigehisa and {Hsieh}, I. -Ta},
        title = "{The Infall Motion in the Low-mass Protostellar Binary NGC 1333 IRAS 4A1/4A2}",
      journal = {apj},
         year = 2019,
        month = nov,
       volume = {885},
       number = {2},
          eid = {98},
        pages = {98},
          doi = {10.3847/1538-4357/ab4818},
archivePrefix = {arXiv},
       eprint = {1909.12443},
 primaryClass = {astro-ph.SR},
       adsurl = {https://ui.adsabs.harvard.edu/abs/2019ApJ...885...98S}
}

@ARTICLE{orti18,
       author = {{Ortiz-Le{\'o}n}, Gisela N. and {Loinard}, Laurent and {Dzib}, Sergio A. and {Galli}, Phillip A.~B. and {Kounkel}, Marina and {Mioduszewski}, Amy J. and {Rodr{\'\i}guez}, Luis F. and {Torres}, Rosa M. and {Hartmann}, Lee and {Boden}, Andrew F. and {Evans}, Neal J., II and {Brice{\~n}o}, Cesar and {Tobin}, John J.},
        title = "{The Gould{\textquoteright}s Belt Distances Survey (GOBELINS). V. Distances and Kinematics of the Perseus Molecular Cloud}",
      journal = {apj},
         year = 2018,
        month = sep,
       volume = {865},
       number = {1},
          eid = {73},
        pages = {73},
          doi = {10.3847/1538-4357/aada49},
archivePrefix = {arXiv},
       eprint = {1808.03499},
 primaryClass = {astro-ph.SR},
       adsurl = {https://ui.adsabs.harvard.edu/abs/2018ApJ...865...73O}
}

@ARTICLE{kris10,
       author = {{Kristensen}, L.~E. and {Visser}, R. and {van Dishoeck}, E.~F. and {Y{\i}ld{\i}z}, U.~A. and {Doty}, S.~D. and {Herczeg}, G.~J. and {Liu}, F. -C. and {Parise}, B. and {J{\o}rgensen}, J.~K. and {van Kempen}, T.~A. and {Brinch}, C. and {Wampfler}, S.~F. and {Bruderer}, S. and {Benz}, A.~O. and {Hogerheijde}, M.~R. and {Deul}, E. and {Bachiller}, R. and {Baudry}, A. and {Benedettini}, M. and {Bergin}, E.~A. and {Bjerkeli}, P. and {Blake}, G.~A. and {Bontemps}, S. and {Braine}, J. and {Caselli}, P. and {Cernicharo}, J. and {Codella}, C. and {Daniel}, F. and {de Graauw}, Th. and {di Giorgio}, A.~M. and {Dominik}, C. and {Encrenaz}, P. and {Fich}, M. and {Fuente}, A. and {Giannini}, T. and {Goicoechea}, J.~R. and {Helmich}, F. and {Herpin}, F. and {Jacq}, T. and {Johnstone}, D. and {Kaufman}, M.~J. and {Larsson}, B. and {Lis}, D. and {Liseau}, R. and {Marseille}, M. and {McCoey}, C. and {Melnick}, G. and {Neufeld}, D. and {Nisini}, B. and {Olberg}, M. and {Pearson}, J.~C. and {Plume}, R. and {Risacher}, C. and {Santiago-Garc{\'\i}a}, J. and {Saraceno}, P. and {Shipman}, R. and {Tafalla}, M. and {Tielens}, A.~G.~G.~M. and {van der Tak}, F. and {Wyrowski}, F. and {Beintema}, D. and {de Jonge}, A. and {Dieleman}, P. and {Ossenkopf}, V. and {Roelfsema}, P. and {Stutzki}, J. and {Whyborn}, N.},
        title = "{Water in low-mass star-forming regions with Herschel . HIFI spectroscopy of NGC 1333}",
      journal = {aap},
         year = 2010,
        month = oct,
       volume = {521},
          eid = {L30},
        pages = {L30},
          doi = {10.1051/0004-6361/201015100},
archivePrefix = {arXiv},
       eprint = {1007.3031},
 primaryClass = {astro-ph.SR},
       adsurl = {https://ui.adsabs.harvard.edu/abs/2010A&A...521L..30K}
}

@ARTICLE{yild10,
       author = {{Y{\i}ld{\i}z}, U.~A. and {van Dishoeck}, E.~F. and {Kristensen}, L.~E. and {Visser}, R. and {J{\o}rgensen}, J.~K. and {Herczeg}, G.~J. and {van Kempen}, T.~A. and {Hogerheijde}, M.~R. and {Doty}, S.~D. and {Benz}, A.~O. and {Bruderer}, S. and {Wampfler}, S.~F. and {Deul}, E. and {Bachiller}, R. and {Baudry}, A. and {Benedettini}, M. and {Bergin}, E. and {Bjerkeli}, P. and {Blake}, G.~A. and {Bontemps}, S. and {Braine}, J. and {Caselli}, P. and {Cernicharo}, J. and {Codella}, C. and {Daniel}, F. and {di Giorgio}, A.~M. and {Dominik}, C. and {Encrenaz}, P. and {Fich}, M. and {Fuente}, A. and {Giannini}, T. and {Goicoechea}, J.~R. and {de Graauw}, Th. and {Helmich}, F. and {Herpin}, F. and {Jacq}, T. and {Johnstone}, D. and {Larsson}, B. and {Lis}, D. and {Liseau}, R. and {Liu}, F. -C. and {Marseille}, M. and {McCoey}, C. and {Melnick}, G. and {Neufeld}, D. and {Nisini}, B. and {Olberg}, M. and {Parise}, B. and {Pearson}, J.~C. and {Plume}, R. and {Risacher}, C. and {Santiago-Garc{\'\i}a}, J. and {Saraceno}, P. and {Shipman}, R. and {Tafalla}, M. and {Tielens}, A.~G.~G.~M. and {van der Tak}, F. and {Wyrowski}, F. and {Dieleman}, P. and {Jellema}, W. and {Ossenkopf}, V. and {Schieder}, R. and {Stutzki}, J.},
        title = "{Herschel/HIFI observations of high-J CO lines in the NGC 1333 low-mass star-forming region}",
      journal = {aap},
         year = 2010,
        month = oct,
       volume = {521},
          eid = {L40},
        pages = {L40},
          doi = {10.1051/0004-6361/201015119},
archivePrefix = {arXiv},
       eprint = {1008.0867},
 primaryClass = {astro-ph.SR},
       adsurl = {https://ui.adsabs.harvard.edu/abs/2010A&A...521L..40Y}
}

@ARTICLE{evan99,
       author = {{Evans}, Neal J., II},
        title = "{Physical Conditions in Regions of Star Formation}",
      journal = {araa},
         year = 1999,
        month = jan,
       volume = {37},
        pages = {311-362},
          doi = {10.1146/annurev.astro.37.1.311},
archivePrefix = {arXiv},
       eprint = {astro-ph/9905050},
 primaryClass = {astro-ph},
       adsurl = {https://ui.adsabs.harvard.edu/abs/1999ARA&A..37..311E}
}

@INPROCEEDINGS{myer00,
       author = {{Myers}, P.~C. and {Evans}, N.~J., II and {Ohashi}, N.},
        title = "{Observations of Infall in Star-Forming Regions}",
    booktitle = {Protostars and Planets IV},
         year = 2000,
       editor = {{Mannings}, V. and {Boss}, A.~P. and {Russell}, S.~S.},
        month = may,
        pages = {217},
       adsurl = {https://ui.adsabs.harvard.edu/abs/2000prpl.conf..217M}
}

@ARTICLE{fran01,
       author = {{Di Francesco}, James and {Myers}, Philip C. and {Wilner}, David J. and {Ohashi}, Nagayoshi and {Mardones}, Diego},
        title = "{Infall, Outflow, Rotation, and Turbulent Motions of Dense Gas within NGC 1333 IRAS 4}",
      journal = {apj},
         year = 2001,
        month = dec,
       volume = {562},
       number = {2},
        pages = {770-789},
          doi = {10.1086/323854},
archivePrefix = {arXiv},
       eprint = {astro-ph/0108022},
 primaryClass = {astro-ph},
       adsurl = {https://ui.adsabs.harvard.edu/abs/2001ApJ...562..770D}
}

@ARTICLE{lyo14,
       author = {{Lyo}, A. -Ran and {Kim}, Jongsoo and {Byun}, Do-Young and {Lee}, Ho-Gyu},
        title = "{Unbiased Water and Methanol Maser Surveys of NGC 1333}",
      journal = {aj},
         year = 2014,
        month = nov,
       volume = {148},
       number = {5},
          eid = {80},
        pages = {80},
          doi = {10.1088/0004-6256/148/5/80},
archivePrefix = {arXiv},
       eprint = {1407.6776},
 primaryClass = {astro-ph.GA},
       adsurl = {https://ui.adsabs.harvard.edu/abs/2014AJ....148...80L}
}

@ARTICLE{guer24,
       author = {{Guerra-Alvarado}, O.~M. and {van der Marel}, N. and {Di Francesco}, J. and {Looney}, L.~W. and {Tobin}, J.~J. and {Cox}, E.~G. and {Sheehan}, P.~D. and {Wilner}, D.~J. and {Mac{\'\i}as}, E. and {Carrasco-Gonz{\'a}lez}, C.},
        title = "{IRAS4A1: Multiwavelength continuum analysis of a very flared Class 0 disk}",
      journal = {aap},
         year = 2024,
        month = jan,
       volume = {681},
          eid = {A82},
        pages = {A82},
          doi = {10.1051/0004-6361/202347685},
archivePrefix = {arXiv},
       eprint = {2310.11999},
 primaryClass = {astro-ph.EP},
       adsurl = {https://ui.adsabs.harvard.edu/abs/2024A&A...681A..82G}
}

@ARTICLE{cout13,
       author = {{Coutens}, A. and {Vastel}, C. and {Cabrit}, S. and {Codella}, C. and {Kristensen}, L.~E. and {Ceccarelli}, C. and {van Dishoeck}, E.~F. and {Boogert}, A.~C.~A. and {Bottinelli}, S. and {Castets}, A. and {Caux}, E. and {Comito}, C. and {Demyk}, K. and {Herpin}, F. and {Lefloch}, B. and {McCoey}, C. and {Mottram}, J.~C. and {Parise}, B. and {Taquet}, V. and {van der Tak}, F.~F.~S. and {Visser}, R. and {Y{\i}ld{\i}z}, U.~A.},
        title = "{Deuterated water in the solar-type protostars NGC 1333 IRAS 4A and IRAS 4B}",
      journal = {aap},
         year = 2013,
        month = dec,
       volume = {560},
          eid = {A39},
        pages = {A39},
          doi = {10.1051/0004-6361/201322400},
archivePrefix = {arXiv},
       eprint = {1310.7365},
 primaryClass = {astro-ph.GA},
       adsurl = {https://ui.adsabs.harvard.edu/abs/2013A&A...560A..39C}
}

@ARTICLE{bott07,
       author = {{Bottinelli}, S. and {Ceccarelli}, C. and {Williams}, J.~P. and {Lefloch}, B.},
        title = "{Hot corinos in NGC 1333-IRAS4B and IRAS2A}",
      journal = {aap},
         year = 2007,
        month = feb,
       volume = {463},
       number = {2},
        pages = {601-610},
          doi = {10.1051/0004-6361:20065139},
archivePrefix = {arXiv},
       eprint = {astro-ph/0611480},
 primaryClass = {astro-ph},
       adsurl = {https://ui.adsabs.harvard.edu/abs/2007A&A...463..601B}
}

@ARTICLE{brin09,
       author = {{Brinch}, C. and {J{\o}rgensen}, J.~K. and {Hogerheijde}, M.~R.},
        title = "{The kinematics of NGC 1333-IRAS2A - a true Class 0 protostar}",
      journal = {aap},
         year = 2009,
        month = jul,
       volume = {502},
       number = {1},
        pages = {199-205},
          doi = {10.1051/0004-6361/200810831},
archivePrefix = {arXiv},
       eprint = {0905.4575},
 primaryClass = {astro-ph.GA},
       adsurl = {https://ui.adsabs.harvard.edu/abs/2009A&A...502..199B}
}

@ARTICLE{tafa15,
       author = {{Tafalla}, M. and {Bachiller}, R. and {Lefloch}, B. and {Rodr{\'\i}guez-Fern{\'a}ndez}, N. and {Codella}, C. and {L{\'o}pez-Sepulcre}, A. and {Podio}, L.},
        title = "{Fast molecular jet from L1157-mm}",
      journal = {aap},
         year = 2015,
        month = jan,
       volume = {573},
          eid = {L2},
        pages = {L2},
          doi = {10.1051/0004-6361/201425255},
archivePrefix = {arXiv},
       eprint = {1412.1476},
 primaryClass = {astro-ph.GA},
       adsurl = {https://ui.adsabs.harvard.edu/abs/2015A&A...573L...2T}
}

@ARTICLE{step13,
       author = {{Stephens}, Ian W. and {Looney}, Leslie W. and {Kwon}, Woojin and {Hull}, Charles L.~H. and {Plambeck}, Richard L. and {Crutcher}, Richard M. and {Chapman}, Nicholas and {Novak}, Giles and {Davidson}, Jacqueline and {Vaillancourt}, John E. and {Shinnaga}, Hiroko and {Matthews}, Tristan},
        title = "{The Magnetic Field Morphology of the Class 0 Protostar L1157-mm}",
      journal = {apjl},
         year = 2013,
        month = may,
       volume = {769},
       number = {1},
          eid = {L15},
        pages = {L15},
          doi = {10.1088/2041-8205/769/1/L15},
archivePrefix = {arXiv},
       eprint = {1304.6739},
 primaryClass = {astro-ph.SR},
       adsurl = {https://ui.adsabs.harvard.edu/abs/2013ApJ...769L..15S}
}

@ARTICLE{bhat22,
       author = {{Bhat}, Bratati and {Gorai}, Prasanta and {Mondal}, Suman Kumar and {Chakrabarti}, Sandip K. and {Das}, Ankan},
        title = "{Radiative transfer modeling of the observed line profiles in G31.41+0.31}",
      journal = {Advances in Space Research},
         year = 2022,
        month = jan,
       volume = {69},
       number = {1},
        pages = {415-437},
          doi = {10.1016/j.asr.2021.07.011},
archivePrefix = {arXiv},
       eprint = {2107.04979},
 primaryClass = {astro-ph.GA},
       adsurl = {https://ui.adsabs.harvard.edu/abs/2022AdSpR..69..415B}
}

@ARTICLE{bhat23,
       author = {{Bhat}, Bratati and {Kar}, Rumela and {Mondal}, Suman Kumar and {Ghosh}, Rana and {Gorai}, Prasanta and {Shimonishi}, Takashi and {Tanaka}, Kei E.~I. and {Furuya}, Kenji and {Das}, Ankan},
        title = "{Chemical Evolution of Some Selected Complex Organic Molecules in Low-mass Star-forming Regions}",
      journal = {The Astrophysical Journal},
         year = 2023,
        month = dec,
       volume = {958},
       number = {2},
          eid = {111},
        pages = {111},
          doi = {10.3847/1538-4357/acfc4d},
archivePrefix = {arXiv},
       eprint = {2308.10211},
 primaryClass = {astro-ph.SR},
       adsurl = {https://ui.adsabs.harvard.edu/abs/2023ApJ...958..111B}
}

@ARTICLE{quit24,
       author = {{Quiti{\'a}n-Lara}, Heidy M. and {Fantuzzi}, Felipe and {Mason}, Nigel J. and {Boechat-Roberty}, Heloisa M.},
        title = "{Decoding the molecular complexity of the solar-type protostar NGC 1333 IRAS 4A}",
      journal = {mnras},
         year = 2024,
        month = feb,
       volume = {527},
       number = {4},
        pages = {10294-10308},
          doi = {10.1093/mnras/stad3873},
       adsurl = {https://ui.adsabs.harvard.edu/abs/2024MNRAS.52710294Q}
}

@ARTICLE{bott04,
       author = {{Bottinelli}, S. and {Ceccarelli}, C. and {Lefloch}, B. and {Williams}, J.~P. and {Castets}, A. and {Caux}, E. and {Cazaux}, S. and {Maret}, S. and {Parise}, B. and {Tielens}, A.~G.~G.~M.},
        title = "{Complex Molecules in the Hot Core of the Low-Mass Protostar NGC 1333 IRAS 4A}",
      journal = {apj},
         year = 2004,
        month = nov,
       volume = {615},
       number = {1},
        pages = {354-358},
          doi = {10.1086/423952},
archivePrefix = {arXiv},
       eprint = {astro-ph/0407154},
 primaryClass = {astro-ph},
       adsurl = {https://ui.adsabs.harvard.edu/abs/2004ApJ...615..354B}
}

@ARTICLE{mare02,
       author = {{Maret}, S. and {Ceccarelli}, C. and {Caux}, E. and {Tielens}, A.~G.~G.~M. and {Castets}, A.},
        title = "{Water emission in NGC 1333-IRAS 4. The physical structure of the envelope}",
      journal = {aap},
         year = 2002,
        month = nov,
       volume = {395},
        pages = {573-585},
          doi = {10.1051/0004-6361:20021334},
archivePrefix = {arXiv},
       eprint = {astro-ph/0209366},
 primaryClass = {astro-ph},
       adsurl = {https://ui.adsabs.harvard.edu/abs/2002A&A...395..573M}
}

@ARTICLE{chah24,
       author = {{Chahine}, Layal and {Ceccarelli}, Cecilia and {De Simone}, Marta and {Chandler}, Claire J. and {Codella}, Claudio and {Podio}, Linda and {L{\'o}pez-Sepulcre}, Ana and {Sakai}, Nami and {Loinard}, Laurent and {Bouvier}, Mathilde and {Caselli}, Paola and {Vastel}, Charlotte and {Bianchi}, Eleonora and {Cuello}, Nicol{\'a}s and {Fontani}, Francesco and {Johnstone}, Doug and {Sabatini}, Giovanni and {Hanawa}, Tomoyuki and {Zhang}, Ziwei E. and {Aikawa}, Yuri and {Busquet}, Gemma and {Caux}, Emmanuel and {Dur{\'a}n}, Aurore and {Herbst}, Eric and {M{\'e}nard}, Fran{\c{c}}ois and {Segura-Cox}, Dominique and {Svoboda}, Brian and {Balucani}, Nadia and {Charnley}, Steven and {Dulieu}, Fran{\c{c}}ois and {Evans}, Lucy and {Fedele}, Davide and {Feng}, Siyi and {Hama}, Tetsuya and {Hirota}, Tomoya and {Isella}, Andrea and {J{\'\i}menez-Serra}, Izaskun and {Lefloch}, Bertrand and {Maud}, Luke T. and {Maureira}, Mar{\'\i}a Jos{\'e} and {Miotello}, Anna and {Moellenbrock}, George and {Nomura}, Hideko and {Oba}, Yasuhiro and {Ohashi}, Satoshi and {Okoda}, Yuki and {Oya}, Yoko and {Pineda}, Jaime and {Rimola}, Albert and {Sakai}, Takeshi and {Shirley}, Yancy and {Testi}, Leonardo and {Viti}, Serena and {Watanabe}, Naoki and {Watanabe}, Yoshimasa and {Zhang}, Yichen and {Yamamoto}, Satoshi},
        title = "{Multiple chemical tracers finally unveil the intricate NGC 1333 IRAS 4A outflow system. FAUST XVI}",
      journal = {mnras},
         year = 2024,
        month = jun,
       volume = {531},
       number = {2},
        pages = {2653-2668},
          doi = {10.1093/mnras/stae1320},
archivePrefix = {arXiv},
       eprint = {2405.12735},
 primaryClass = {astro-ph.GA},
       adsurl = {https://ui.adsabs.harvard.edu/abs/2024MNRAS.531.2653C}
}

@ARTICLE{wies13,
       author = {{Wiesenfeld}, L. and {Faure}, A.},
        title = "{Rotational quenching of H$_{2}$CO by molecular hydrogen: cross-sections, rates and pressure broadening}",
      journal = {mnras},
         year = 2013,
        month = jul,
       volume = {432},
       number = {3},
        pages = {2573-2578},
          doi = {10.1093/mnras/stt616},
archivePrefix = {arXiv},
       eprint = {1304.4804},
 primaryClass = {astro-ph.SR},
       adsurl = {https://ui.adsabs.harvard.edu/abs/2013MNRAS.432.2573W}
}

@ARTICLE{flow06,
       author = {{Flower}, D.~R. and {Pineau Des For{\^e}ts}, G. and {Walmsley}, C.~M.},
        title = "{The importance of the ortho:para H$_{2}$ ratio for the deuteration of molecules during pre-protostellar collapse}",
      journal = {aap},
         year = 2006,
        month = apr,
       volume = {449},
       number = {2},
        pages = {621-629},
          doi = {10.1051/0004-6361:20054246},
archivePrefix = {arXiv},
       eprint = {astro-ph/0601429},
 primaryClass = {astro-ph},
       adsurl = {https://ui.adsabs.harvard.edu/abs/2006A&A...449..621F}
}

@ARTICLE{ivez97,
       author = {{Ivezic}, Zeljko and {Elitzur}, Moshe},
        title = "{Self-similarity and scaling behaviour of infrared emission from radiatively heated dust - I. Theory}",
      journal = {mnras},
         year = 1997,
        month = jun,
       volume = {287},
       number = {4},
        pages = {799-811},
          doi = {10.1093/mnras/287.4.799},
archivePrefix = {arXiv},
       eprint = {astro-ph/9612164},
 primaryClass = {astro-ph},
       adsurl = {https://ui.adsabs.harvard.edu/abs/1997MNRAS.287..799I}
}

@ARTICLE{paga09,
       author = {{Pagani}, L. and {Vastel}, C. and {Hugo}, E. and {Kokoouline}, V. and {Greene}, C.~H. and {Bacmann}, A. and {Bayet}, E. and {Ceccarelli}, C. and {Peng}, R. and {Schlemmer}, S.},
        title = "{Chemical modeling of <ASTROBJ>L183</ASTROBJ> (<ASTROBJ>L134N</ASTROBJ>): an estimate of the ortho/para H\{\_2\} ratio}",
      journal = {aap},
         year = 2009,
        month = feb,
       volume = {494},
       number = {2},
        pages = {623-636},
          doi = {10.1051/0004-6361:200810587},
archivePrefix = {arXiv},
       eprint = {0810.1861},
 primaryClass = {astro-ph},
       adsurl = {https://ui.adsabs.harvard.edu/abs/2009A&A...494..623P}
}

@ARTICLE{vank08,
       author = {{van Kempen}, T.~A. and {Doty}, S.~D. and {van Dishoeck}, E.~F. and {Hogerheijde}, M.~R. and {J{\o}rgensen}, J.~K.},
        title = "{Modeling water emission from low-mass protostellar envelopes}",
      journal = {aap},
         year = 2008,
        month = sep,
       volume = {487},
       number = {3},
        pages = {975-991},
          doi = {10.1051/0004-6361:200809426},
archivePrefix = {arXiv},
       eprint = {0805.0772},
 primaryClass = {astro-ph},
       adsurl = {https://ui.adsabs.harvard.edu/abs/2008A&A...487..975V}
}

@ARTICLE{jorg02,
       author = {{J{\o}rgensen}, J.~K. and {Sch{\"o}ier}, F.~L. and {van Dishoeck}, E.~F.},
        title = "{Physical structure and CO abundance of low-mass protostellar envelopes}",
      journal = {aap},
         year = 2002,
        month = jul,
       volume = {389},
        pages = {908-930},
          doi = {10.1051/0004-6361:20020681},
archivePrefix = {arXiv},
       eprint = {astro-ph/0205068},
 primaryClass = {astro-ph},
       adsurl = {https://ui.adsabs.harvard.edu/abs/2002A&A...389..908J}
}

@ARTICLE{das24,
       author = {{Das}, Ankan},
        title = "{Astrochemistry: The study of chemical processes in space}",
      journal = {Life Sciences in Space Research},
         year = 2024,
        month = nov,
       volume = {43},
        pages = {43-53},
          doi = {10.1016/j.lssr.2024.10.005},
       adsurl = {https://ui.adsabs.harvard.edu/abs/2024LSSR...43...43D}
}

@ARTICLE{gora20,
       author = {{Gorai}, Prasanta and {Bhat}, Bratati and {Sil}, Milan and {Mondal}, Suman K. and {Ghosh}, Rana and {Chakrabarti}, Sandip K. and {Das}, Ankan},
        title = "{Identification of Prebiotic Molecules Containing Peptide-like Bonds in a Hot Molecular Core, G10.47+0.03}",
      journal = {apj},
         year = 2020,
        month = jun,
       volume = {895},
       number = {2},
          eid = {86},
        pages = {86},
          doi = {10.3847/1538-4357/ab8871},
archivePrefix = {arXiv},
       eprint = {2003.09188},
 primaryClass = {astro-ph.SR},
       adsurl = {https://ui.adsabs.harvard.edu/abs/2020ApJ...895...86G}
}

@ARTICLE{baek22,
       author = {{Baek}, Giseon and {Lee}, Jeong-Eun and {Hirota}, Tomoya and {Kim}, Kee-Tae and {Kim}, Mi Kyoung},
        title = "{Complex Organic Molecules Detected in 12 High-mass Star-forming Regions with Atacama Large Millimeter/submillimeter Array}",
      journal = {apj},
         year = 2022,
        month = nov,
       volume = {939},
       number = {2},
          eid = {84},
        pages = {84},
          doi = {10.3847/1538-4357/ac81d3},
archivePrefix = {arXiv},
       eprint = {2207.08223},
 primaryClass = {astro-ph.GA},
       adsurl = {https://ui.adsabs.harvard.edu/abs/2022ApJ...939...84B}
}

@ARTICLE{mond23,
       author = {{Mondal}, Suman Kumar and {Iqbal}, Wasim and {Gorai}, Prasanta and {Bhat}, Bratati and {Wakelam}, Valentine and {Das}, Ankan},
        title = "{Investigating the hot molecular core, G10.47+0.03: A pit of nitrogen-bearing complex organic molecules}",
      journal = {aap},
         year = 2023,
        month = jan,
       volume = {669},
          eid = {A71},
        pages = {A71},
          doi = {10.1051/0004-6361/202243802},
archivePrefix = {arXiv},
       eprint = {2211.03066},
 primaryClass = {astro-ph.GA},
       adsurl = {https://ui.adsabs.harvard.edu/abs/2023A&A...669A..71M}
}

@ARTICLE{shim20,
       author = {{Shimonishi}, Takashi and {Das}, Ankan and {Sakai}, Nami and {Tanaka}, Kei E.~I. and {Aikawa}, Yuri and {Onaka}, Takashi and {Watanabe}, Yoshimasa and {Nishimura}, Yuri},
        title = "{Chemistry and Physics of a Low-metallicity Hot Core in the Large Magellanic Cloud}",
      journal = {apj},
         year = 2020,
        month = mar,
       volume = {891},
       number = {2},
          eid = {164},
        pages = {164},
          doi = {10.3847/1538-4357/ab6e6b},
archivePrefix = {arXiv},
       eprint = {2001.06982},
 primaryClass = {astro-ph.GA},
       adsurl = {https://ui.adsabs.harvard.edu/abs/2020ApJ...891..164S}
}


\end{document}